\documentclass[11pt]{article}
\pdfoutput=1
\usepackage{jcapmod}

\usepackage{amsmath, amsfonts}
\usepackage[english]{babel}
\usepackage{graphicx}
\usepackage{subcaption}
\usepackage{hyperref}
\usepackage[range-units=brackets, range-phrase={,}, per-mode=reciprocal, mode=math]{siunitx}[=v2]
\usepackage[babel=true]{microtype}

\numberwithin{equation}{section}
\allowdisplaybreaks[1]

\def\d{\mathrm{d}}
\def\ee{\mathrm{e}}
\def\ii{\mathrm{i}}
\def\Neff{N_\mathrm{eff}}
\def\Mpl{M_\mathrm{pl}}
\def\L{\mathcal{L}}
\def\M{\mathcal{M}}
\def\p{{\boldsymbol{p}}}
\def\zhat{{\boldsymbol{\hat{z}}}}

\DeclareRobustCommand{\SkipTocEntry}[4]{}

\begin{document}

\pagenumbering{roman}
\begin{titlepage}
	\baselineskip=15.5pt \thispagestyle{empty}
	
	\bigskip\
	
	\vspace{1cm}
	\begin{center}
		{\fontsize{20.74}{24}\selectfont \sffamily \bfseries Cosmological Implications of\\[8pt]Axion-Matter Couplings}
	\end{center}
	
	\vspace{0.2cm}
	\begin{center}
		{\fontsize{12}{30}\selectfont Daniel Green,$^{\bigstar}$, Yi Guo$^{\bigstar}$ and Benjamin Wallisch$^{\bigstar,\spadesuit}$}
	\end{center}
	
	\begin{center}
		\vskip8pt
		\textsl{$^\bigstar$ Department of Physics, University of California San Diego, La Jolla, CA 92093, USA}
		
		\vskip8pt
		\textsl{$^\spadesuit$ School of Natural Sciences, Institute for Advanced Study, Princeton, NJ 08540, USA}
	\end{center}

	\vspace{1.2cm}
	\hrule \vspace{0.3cm}
	\noindent {\sffamily \bfseries Abstract}\\[0.1cm]
Axions and other light particles appear ubiquitously in physics beyond the Standard Model, with a variety of possible couplings to ordinary matter. Cosmology offers a unique probe of these particles as they can thermalize in the hot environment of the early universe for any such coupling. For sub-MeV particles, their entropy must leave a measurable cosmological signal, usually via the effective number of relativistic particles,~$\Neff$. In this paper, we will revisit the cosmological constraints on the couplings of axions and other pseudo-Nambu-Goldstone bosons to Standard Model fermions from thermalization below the electroweak scale, where these couplings are marginal and give contributions to the radiation density of $\Delta\Neff > 0.027$. We update the calculation of the production rates to eliminate unnecessary approximations and find that the cosmological bounds on these interactions are complementary to astrophysical constraints, e.g.~from supernova~SN~1987A. We additionally provide quantitative explanations for these bounds and their relationship.
	\vskip10pt
	\hrule
	\vskip10pt
\end{titlepage}

\thispagestyle{empty}
\setcounter{page}{2}
\tableofcontents

\clearpage
\pagenumbering{arabic}
\setcounter{page}{1}

\clearpage
\section{Introduction}
\label{sec:introduction}

Light particles with very weak couplings to the Standard Model are highly-motivated experimental targets from a number of perspectives. Axions and axion-like particles have been proposed as solutions to fine-tuning problems like the strong CP~problem~\cite{Peccei:1977hh, Weinberg:1977ma, Hook:2018dlk} and the hierarchy problem~\cite{Graham:2015cka}. Furthermore, a cold component of the axion may form a viable dark matter candidate~\cite{Preskill:1982cy, Abbott:1982af, Dine:1982ah}. Alternatively, light particles may take the form of Goldstone or pseudo-Goldstone modes that arise as a consequence of symmetry breaking, including in models of flavor~\cite{Davidson:1981zd, Wilczek:1982rv, Reiss:1982sq, Feng:1997tn}~(familons), neutrino masses~\cite{Chikashige:1980ui, Chacko:2003dt}~(majorons) and supersymmetry~(gravitino). More broadly, top-down models suggest that there could be a number of additional sectors with very weak or gravitational couplings to the Standard Model~\cite{Svrcek:2006yi, Arvanitaki:2009fg, Marsh:2015xka, Arkani-Hamed:2016rle, Chacko:2016hvu, Chacko:2018vss}. The plethora of possibilities for light particles~\cite{Jaeckel:2010ni, Essig:2013lka} is mirrored in the variety of dark matter candidates and dark sectors that are being actively explored~\cite{Graham:2015ouw}.\medskip

Cosmology plays a vital role in our investigations of this vast landscape. While the list of possible couplings to the Standard Model~(SM) are enormous, if any of them is sufficient to thermalize one of these particles, its relic energy density is detectable through its gravitational influence~\cite{Braaten:1991dd, Bolz:2000fu, Masso:2002np, Graf:2010tv, Cadamuro:2010cz, Cadamuro:2011fd, Brust:2013ova, Weinberg:2013kea, Salvio:2013iaa, Baumann:2016wac}. Thermalized particles with masses $m \ll \SI{1}{eV}$ are relativistic during the radiation era and the cosmological constraints can be inferred from the effective number of relativistic species~$\Neff$. Heavier particles with $m < \SI{100}{eV}$ will contribute to the sum of neutrino masses and are also constrained by the evolution of the universe at low redshifts. Most importantly, when its mass is sub-\si{MeV}, the entropy carried by a new particle cannot be eliminated by its decay or annihilation and, therefore, always leaves a cosmological signal in some combination of relic abundances (big bang nucleosynthesis/BBN)~\cite{Cyburt:2015mya, Pitrou:2018cgg, Fields:2019pfx}, cosmic microwave background~(CMB)~\cite{Bashinsky:2003tk, Follin:2015hya, Baumann:2015rya, Brust:2017nmv, Choi:2018gho, Planck:2018vyg, Blinov:2020hmc} and/or large-scale structure~\cite{Baumann:2017lmt, Baumann:2017gkg, Baumann:2019keh, Green:2020fjb} observables.\medskip

The majority of couplings of light particles to the Standard Model are irrelevant. The combination of symmetries needed to protect the mass of the particle combined with the limited set of gauge-invariant operators in the Standard Model usually ensures that they are dimension five or larger~\cite{Jaeckel:2010ni, Essig:2013lka, Brust:2013ova}. As a result, thermalization usually occurs at high temperatures where all the particles involved are relativistic. As a result, the strongest constraints on these couplings will arise from thermalization well above the electroweak scale. This implies that their contributions to~$\Neff$ are diluted to their minimum value, $\Delta\Neff = 0.027\,g_s$, where~$g_s$ is the number of degrees of freedom of the light relic (see e.g.~\cite{Baumann:2018muz, Wallisch:2018rzj, Green:2019glg} for reviews).\medskip

Couplings of axions (familons) to matter are an exception to this general pattern.\footnote{For simplicity, we will refer to all scalar particles interacting with matter as axions, whether we assume specific models or independent couplings, despite the fact that the latter might be more naturally referred to as familons in most of our cases or, more generally, pseudo-Nambu-Goldstone bosons~(pNGBs).} Starting from a manifestly shift-symmetric form, a scalar field~$\phi$ can be coupled to the SM~fermions~$\psi_i$ via
\begin{equation}
	\begin{split}
		\mathcal{L}_{\phi \psi}	&=-\frac{\partial_{\mu} \phi}{\Lambda_\psi} \bar{\psi}_{i} \gamma^{\mu}\left(g_{V}^{i j}+g_{A}^{i j} \gamma^{5}\right) \psi_{j} \\
								& \rightarrow \frac{\phi}{\Lambda_\psi}\left(\mathrm{i} H \bar{\psi}_{L, i}\left[\left(\lambda_{i}-\lambda_{j}\right) g_{V}^{i j}+\left(\lambda_{i}+\lambda_{j}\right) g_{A}^{i j}\right] \psi_{R, j}+\mathrm{h.c.}\right)+\mathcal{O}\left(\phi^{2}\right) ,
	\end{split} \label{eq:lagrangian}
\end{equation}
where we integrated by parts and used the equations of motion with the Higgs doublet~$H$, the left-/right-handed spinors~$\psi_{L,R} \equiv \frac{1}{2}\left(1 \mp \gamma^{5}\right) \psi$, the Yukawa couplings~$\lambda_i \equiv \sqrt{2} m_i/v$ and the Higgs vacuum expectation value~$v = \SI{246}{GeV}$.\footnote{For simplicity, we suppressed the $SU(2)_L$ and $SU(3)_c$~structures which take the same form as for the SM~Yukawa couplings~\cite{Peskin:1995ev}.} Importantly, we see on the second line that this interaction is effectively dimension four in the presence of a non-zero Higgs vacuum expectation value. By dimensional analysis, the production rate of~$\phi$ is proportional to temperature (since the effective coupling is dimensionless) and, therefore, exceeds the Hubble rate at low temperatures, which scales as temperature squared during radiation domination. However, this argument is only true above the mass of the SM~fermion since the production rate will again become negligibly small once the number density of the fermion is sufficiently Boltzmann suppressed. This possibility is particularly intriguing because the decoupling temperature is effectively below the mass of the associated fermion and, therefore, gives a larger contribution to~$\Neff$ which is more easily constrained or detected with near-term CMB experiments, such as Simons Observatory~(SO) and~\mbox{CMB-S4}~\cite{SimonsObservatory:2018koc, Abazajian:2019eic}.\medskip

For the above reasons, cosmological constraints on axion-matter couplings have received significant attention in the literature~\cite{Brust:2013ova, Baumann:2016wac, Ferreira:2018vjj, DEramo:2018vss, Arias-Aragon:2020shv, Ghosh:2020vti, Ferreira:2020bpb, Dror:2021nyr}. Yet, while the origin of the constraint is straightforward to estimate qualitatively, precise numerical bounds depend on a number of details that have only been partially explored in the literature. Most essentially, the thermal production rate and/or decoupling calculations are often approximated in various ways. In some cases, the estimated bounds are substantially stronger than the true bounds~\cite{Baumann:2016wac}.\medskip

The result of our analysis is a more precise calculation of~$\Delta\Neff$ for interactions with charged leptons and heavy quarks, namely the electron, muon and tau~lepton, and the charm, bottom and top~quarks, as a function of each of their coupling strengths. In addition, we provide an intuitive (semi-)analytic explanation for the form of each of these curves. From these results, it is possible to straightforwardly derive the constraints on axion-matter couplings from a given measurement of~$\Neff$. Moreover, we compare current and future cosmological constraints to existing bounds in the literature. Of particular interest is the relation to constraints from~SN~1987A which have recently been inferred for couplings to muons~\cite{Bollig:2020xdr, Croon:2020lrf, Caputo:2021rux}. As both the cosmological and astrophysical bounds are derived from thermal production of the axion, we explore the precise relationship between these bounds in that context.\bigskip

This paper is organized as follows: In Section~\ref{sec:review}, we summarize the cosmology and particle physics of axions, or more generally~pNGBs, and their coupling to SM~fermions studied in this work. This includes a qualitative explanation of the freeze-in (low-temperature re-thermalization and decoupling) and freeze-out (high-temperature decoupling) scenarios that are possible below and above the electroweak symmetry breaking scale, and how they allow measurements of~$\Neff$ to be translated into bounds on these interactions. In Section~\ref{sec:calculations}, we present the calculation of the axion production rates and the coupling constraints. Moreover, we infer the bounds from current and future measurements of~$\Neff$, and describe the physics underlying these cosmological constraints in detail. In Section~\ref{sec:comparison}, we explore the relation between these bounds and those from astrophysical measurements, in particular the cooling rate of~SN~1987A, and quantitatively compare our $\Neff$-based constraints to those and other existing bounds. In Section~\ref{sec:conclusions}, we present our conclusions. A set of appendices contains technical details on the calculation of the axion production rate~(Appendix~\ref{app:computationalDetails}), and the implications of quantum statistics and the presence of the QCD~phase transition on our results~(Appendix~\ref{app:rateComparisonsUncertainties}).

\section{Review of Axions and Axion Cosmology}
\label{sec:review}

Axion-like particles and other pseudo-Nambu-Goldstone bosons arise in a variety of forms, depending on the ultraviolet completion. String theory famously contains a plenitude of axions. Alternatively, the strong CP~problem suggests the need for an axion~$\phi$ with a coupling $\phi \tilde{G}_{\mu\nu} G^{\mu\nu}$, where $G_{\mu\nu}$ is the gluon field strength tensor and $\tilde{G}_{\mu\nu}$ its dual. The key feature of the coupling of such particles is that they preserve a shift symmetry, $\phi \to \phi+c$, with constant~$c$, such that they can be naturally light.\medskip

While the coupling of axions to gauge bosons is often what distinguishes axions from other naturally light scalars, the coupling to matter can and will arise for all such particles. In some cases, the particles are given alternate names such as \textit{familons}. Nevertheless, given the shift symmetry, the leading interactions with the SM~fermions is given by\hskip1pt\footnote{In this paper, we follow the notation of~\cite{Baumann:2016wac} and parameterize the dimensionful axion couplings in terms of the effective mass scale $\Lambda_{ij} \equiv {\Lambda_\psi/[(g_V^{ij})^2 + (g_A^{ij})^2}]^{1/2}$. This parametrization can be straightforwardly converted to other commonly employed notations, such as the inverse scale $\tilde{g}_{ij} = 1/\Lambda_{ij}$, the dimensionless coupling constant $\tilde{\epsilon}_i = 2 m_i/\Lambda_{ii}$ or the decay constant $f_a = \Lambda_\psi$.}
\begin{equation}
	\mathcal{L}_{\phi\psi} = -\frac{\partial_\mu \phi}{\Lambda_\psi} \left(g_V^{ij} J_V^{ij} + g_A^{ij} J_A^{ij}\right) = -\frac{\partial_\mu \phi}{\Lambda_\psi} \bar\psi_{i} \gamma^\mu \left( g_V^{ij} + g_A^{ij} \gamma^{5} \right) \psi_j\, ,
\end{equation}
where the couplings to the vector and axial-vector currents~$J_{V,A}$ are denoted by the subscripts~$V$ and~$A$, respectively. The diagonal vector couplings, $i=j$, vanish due to vector current conservation, i.e.\ diagonal couplings are only present for the axial part. This can also be seen explicitly after integrating by parts and using the equations of motions, as performed in the second line of~\eqref{eq:lagrangian}. While a linear combination of the axial couplings is equivalent to the coupling of axions to gauge bosons due to the chiral anomaly, we only consider the effects of the couplings to matter with no contribution from anomalies in this paper.\medskip

In the cosmological context, the impact of an axion-matter interaction is qualitatively different before and after the electroweak phase transition. Prior to the electroweak phase transition, these couplings are described by dimension-5 operators and the axion interaction rate with SM~particles therefore scales as $\Gamma_\phi \sim T^3/\Lambda_\psi^2$. Meanwhile, the expansion rate of the universe scales as $H \sim T^2/\Mpl$ at those early times. This implies that the axion may be in thermal equilibrium with the rest of the~SM for $T > T_F$, where freeze out at temperature~$T_F$ is defined by $H(T_F) \simeq \Gamma_\phi(T_F)$ (assuming~$T_F$ is above the electroweak scale). For $T < T_F$, axion production becomes inefficient and the axions decouple from the Standard Model. Note that a population of hot axions may exist whether or not axions form the dark matter and are therefore complimentary to many of the direct detection strategies~\cite{Graham:2015ouw}.

\begin{figure}
	\centering
	\includegraphics{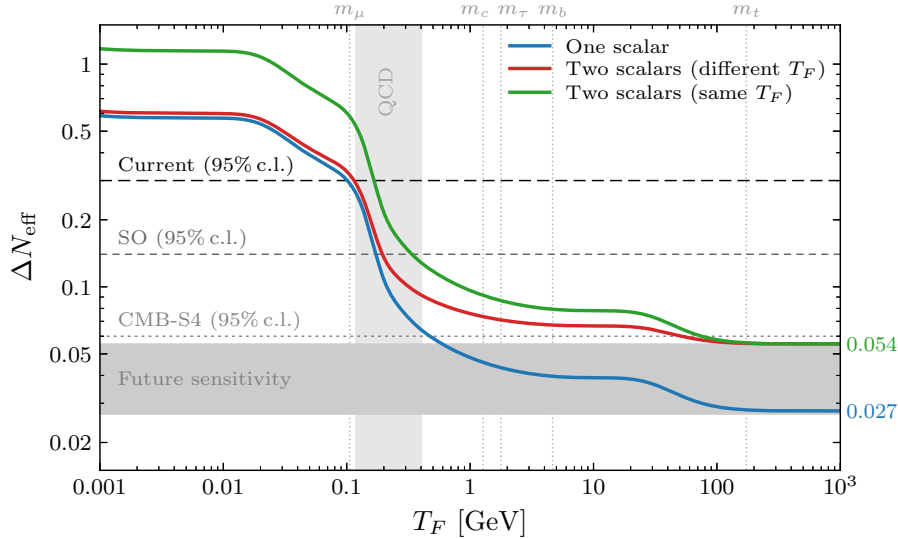}
	\caption{Contribution to~$\Delta\Neff$ from a light particle that decoupled from the Standard Model at a freeze-out temperature~$T_F$. The blue line indicates the contribution for a single real degree of freedom, such as an axion or Goldstone boson. The green and red lines show~$\Delta\Neff(T_F)$ for the case of two scalar degrees of freedom, either with the same decoupling temperature, $T_F \equiv T_{F,1} = T_{F,2},$ or two different decoupling temperatures, $T_F \equiv T_{F,1} \lesssim \SI{e3}{GeV} < T_{F,2}$. The dashed lines indicate the current bound on~$\Delta\Neff$ at 95\%~c.l.\ from Planck~2018 and BAO~data~\cite{Planck:2018vyg}, and the forecasted constraints from the Simons Observatory~(SO)~\cite{SimonsObservatory:2018koc} and \mbox{CMB-S4}~\cite{Abazajian:2019eic}. The gray band illustrates the future sensitivity that might potentially be achieved with a combination of cosmological surveys of the~CMB and large-scale structure, such as CMB-HD~\cite{Sehgal:2019ewc}, MegaMapper~\cite{Schlegel:2019eqc} and PUMA~\cite{PUMA:2019jwd}, cf.~\cite{Baumann:2017gkg, Sailer:2021yzm}. We refer to~\cite{Wallisch:2018rzj, Green:2019glg} for additional details.}
	\label{fig:DeltaNeff_freezeout}
\end{figure}

This description of decoupling is applicable to any particle coupled to the Standard Model through an irrelevant operator. This is the common origin of most (light) thermal relics in the early universe and leads to the standard contribution to the energy density in free-streaming radiation as parameterized by
\begin{equation}
	\Neff = \frac{8}{7} \left(\frac{11}{4}\right)^{\!4/3} \frac{\rho_{\nu}+\rho_\phi}{\rho_{\gamma}} \qquad \to \qquad \Delta\Neff = \frac{8}{7} \left(\frac{11}{4}\right)^{\!4/3} \frac{\rho_\phi}{\rho_{\gamma}}\, ,	\label{eq:DeltaNeff}
\end{equation}
where $\rho_\nu$ and~$\rho_\phi$ are the energy density in neutrinos and axions (or any other light thermal relic beyond the Standard Model), respectively. Given their sub-\si{eV} masses, both are relativistic prior to recombination, which means that $\Neff$~does not distinguish between axions, neutrinos or any other decoupling relativistic species. While $\Neff = 3.044$ in the Standard Model due to the three neutrinos~\cite{Akita:2020szl, Froustey:2020mcq, Bennett:2020zkv}, a thermalized axion or other~pNGB will contribute $\Delta\Neff \geq 0.027$, with this bound being saturated for decoupling above all SM~mass thresholds. For general~$T_F$ and $g_s$~internal degrees of freedom, the contribution is given by\hskip1pt\footnote{We make the assumption that there are no large sources of entropy beyond the Standard Model particles at or below the freeze-out temperature~$T_F$. See e.g.~\cite{Wallisch:2018rzj} for a more detailed discussion.}
\begin{equation}
	\Delta\Neff = g_s \left(\frac{43 / 4}{g_*(T_F)}\right)^{\!4/3} ,
\end{equation}
where~$g_*(T)$ is the effective number of SM~degrees of freedom at temperature~$T$. The effect of lower~$T_F$ is to increase~$\Delta\Neff$ by reducing the amount of entropy converted to photons after decoupling of the axion. This increase is shown in Fig.~\ref{fig:DeltaNeff_freezeout}%
since the SM~particles become massive, annihilate and deposit their energy (and entropy) in the remaining thermal SM~bath. This therefore provides a natural observational target (see also~\cite{Wallisch:2018rzj, Green:2019glg}, for instance).

The absence of a detection with an exclusion of $\Delta\Neff = 0.027$ using future cosmological data would put strong constraints on the coupling strength~$\Lambda_{ij}$~\cite{Baumann:2016wac} since this would exclude the presence of any thermalized relics above the electroweak scale. In this case, no additional light particles could have been in thermal equilibrium with the~SM at any point in the history of the universe (including axions), back to the era of reheating at temperature~$T_R$. Since this requires the would-be freeze-out temperature to be above the reheating temperature, $T_F(\Lambda_\psi) > T_R$, a significant exclusion of $\Delta\Neff = 0.027$ would imply very strong constraints on the axion couplings. If we define $\Lambda^{(\psi)}_{F}(T)$ as the coupling~$\Lambda_\psi$ such that $\Gamma(T) = H(T)$ at temperature~$T$, then our approximate bound would be $\Lambda_\psi \gtrsim \Lambda^{(\psi)}_{F}(T_R)$.\footnote{This also assumes no dramatic increase in the number of degrees of freedom in the Standard Model or non-equilibrium evolution that could further dilute $\Delta\Neff < 0.027$ (see~\cite{Wallisch:2018rzj} for more details and discussion).} For the scales in~\eqref{eq:lagrangian}, this constraint would imply~\cite{Baumann:2016wac}\hskip1pt\footnote{We directly use the results of~\cite{Baumann:2016wac} and do not include any improved calculations of the axion production rate here because the sensitivity to the (unknown) reheating temperature limits the need for a more precise calculation at this point.}
\begin{equation}
	\Lambda_{ij}^{I} \ >\ \left\{
		\begin{array}{ll}
			\displaystyle \SI{1.0e11}{GeV} \ \frac{m_i \mp m_j}{m_\tau} \left(\frac{T_R}{\SI{e10}{GeV}}\right)^{\!1/2}	& \quad	i,j = \text{leptons},	\\[10pt]
			\displaystyle \SI{1.8e13}{GeV} \ \frac{m_i \mp m_j}{m_t} \left(\frac{T_R}{\SI{e10}{GeV}}\right)^{\!1/2}		& \quad i,j = \text{quarks},
		\end{array}
	\right.
\end{equation}
where~$m_i$ are the SM~fermion masses, with $m_\tau \approx \SI{1.8}{GeV}$ and $m_t \approx \SI{173}{GeV}$. We refer to~\cite{Baumann:2016wac} for additional details, including a comparison of the current experimental and the prospective cosmological constraints which will likely be stronger by orders of magnitude except for most interactions involving electrons.\medskip

However, after the electroweak phase transition, there exists a second scenario which is unique to the couplings of SM~fermions to axions (and other~pNGBs). The out-of-equilibrium axions may re-equilibrate and thermalize with the Standard Model after the Higgs acquired its non-zero vacuum expectation value. In this case, the Lagrangian~\eqref{eq:lagrangian} becomes
\begin{equation}
	\L_{\phi \psi} = \ii \frac{\phi}{\Lambda_\psi} \bar \psi_i \left[(m_i-m_j) g_V^{ij} +(m_i+m_j) g_A^{ij} \gamma^5 \right] \psi_j\, ,	\label{eq:lagrangian2}
\end{equation}
which is effectively a dimension-four interaction. This implies that the interaction rate now scales as $\Gamma_\phi \sim m_\psi^2\, T/\Lambda_\psi^2$, which is a weaker temperature dependence than that of the expansion rate, $H \sim T^2$. Depending on the interaction strength, the axions will eventually thermalize and decouple again at later times leaving a much larger contribution to the radiation density as displayed in Fig.~\ref{fig:DeltaNeff_freezeout}. Such a contribution to~$\Delta\Neff$ may already be ruled out with current cosmological datasets or ruled out in the (near) future. Preventing this re-thermalization of the axion to lead to a large axion density and violation of existing (or near-future) constraints on~$\Delta\Neff$ requires the re-equilibration temperature to be smaller than the mass of the respective fermion(s) since the interaction rate~$\Gamma$ becomes Boltzmann suppressed in this regime. This in turn suggests that we can put limits on the axion couplings by effectively trading the reheating temperature~$T_R$ with the fermion mass~$m_\psi$, $\Lambda_\psi \gtrsim \Lambda_F^{(\psi)}(m_\psi)$~\cite{Baumann:2016wac}.

The resulting bounds are typically weaker than those derived from freeze-out above the electroweak scale, as the higher temperatures ultimately lead to more efficient production. Nevertheless, high-temperature freeze-out is more\hskip1pt\footnote{The freeze-in scenario (re-thermalization and decoupling) only assumes that the reheating temperature is larger than the fermion mass(es), \mbox{$T_R > m_\psi$}.} sensitive to assumptions on the reheating temperature and particle content of the universe. Moreover, these bounds will be easier (and therefore earlier) to achieve since the larger contributions to~$\Delta\Neff$ that are generated by the freeze-in process can be more easily measured (or excluded) by a realistic cosmological survey. While order-of-magnitude estimates of the resulting bounds were provided in~\cite{Baumann:2016wac}, the contributions to~$\Delta\Neff$ are at the threshold of current and future CMB~experiments and, therefore, demand a more careful treatment. Given the highly nonlinear relationship between~$\Delta\Neff$ and the fundamental parameters of the model, even seemingly small effects can translate into large differences in the inference of the axion-matter couplings (see e.g.~\textsection\ref{sec:physics} and Appendix~\ref{app:rateComparisonsUncertainties}).

\section{Production Rates and Cosmic Constraints}
\label{sec:calculations}

Relating $\Neff$~measurements to specific models is dependent on a reliable calculation of the production rate. The existence of a constraint usually follows from dimensional analysis, but we also have to put the cosmological constraints into the broader context of experimental probes of axions and other~pNGBs. Calculating the axion production rate accurately is difficult as some approximations are unreliable when $T \approx m_\psi$~\cite{Baumann:2016wac}.\footnote{Prior calculations of the axion production rate have assumed relativistic particles and a high-temperature limit~\cite{Baumann:2016wac} or used Boltzmann statistics ignoring Bose enhancement and Pauli blocking~\cite{DEramo:2018vss}, for instance.} In this section, we will therefore recalculate these rates for the couplings of axions to SM~matter particles without these approximations~(\textsection\ref{sec:productionRates}). We will then derive predictions of~$\Delta\Neff$ and observational bounds on these interactions, focusing on the diagonal couplings, $\Lambda_i \equiv \Lambda_\psi/g_A^{ii}$, for simplicity~(\textsection\ref{sec:constraints}). Moreover, we will provide a detailed discussion of the physics underlying these constraints~(\textsection\ref{sec:physics}).

\subsection{Computation of Production Rates}
\label{sec:productionRates}

In the following, we summarize the calculation of the interaction rate of axions and other pseudo-Nambu-Goldstone bosons with SM~fermions. While we will put an emphasis here on the conceptual steps and relevant physical processes, we refer to Appendix~\ref{app:computationalDetails} for the technical details.\medskip

The leading processes of producing axions in the early universe with diagonal interactions described by~\eqref{eq:lagrangian2} after electroweak symmetry breaking are shown in Fig.~\ref{fig:FeynmanDiagrams}: %
\begin{figure}
	\centering
	\begin{subfigure}[b]{0.33\textwidth}
		\centering
		\includegraphics{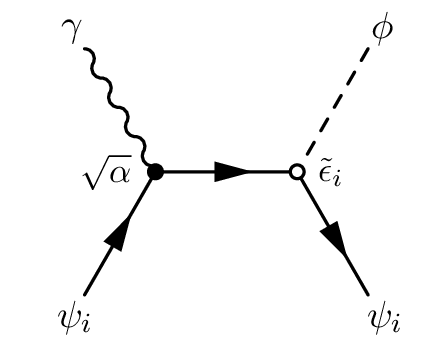}
		\caption{Compton-like scattering.}
	\end{subfigure}
	\hspace{1.cm}
	\begin{subfigure}[b]{0.33\textwidth}
		\centering
		\includegraphics{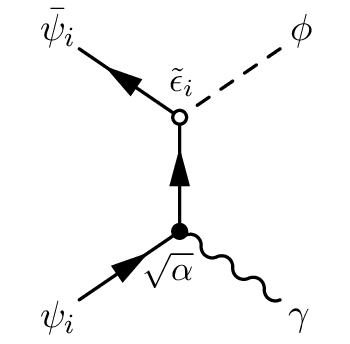}
		\caption{Fermion annihilation.}
	\end{subfigure}
	\caption{Feynman diagrams for the dominant production channels of axions and other pseudo-Nambu-Goldstone bosons via the coupling to charged fermions below the electroweak scale: (a)~Compton-like scattering and (b)~fermion annihilation. For quarks, the coupling to photons is replaced by that to gluons. In addition to the displayed $s$- and $t$-channel diagrams, there are $u$-channel diagrams which are not shown.}
	\label{fig:FeynmanDiagrams}
\end{figure}
(a)~Compton-like scattering, $\psi_i + \{\gamma,g\} \to \psi_i + \phi$, and (b)~fermion-antifermion annihilation, $\psi_i + \bar{\psi_i} \to \phi + \{\gamma,g\}$, where we denoted the photon and gluon by~$\gamma$ and~$g$, respectively. The scattering amplitudes of these production channels are given by~\cite{Baumann:2016wac}
\begin{equation}
	\sum |\M|^2_{(a)} = 16\pi \hskip1pt A_\psi \, |\tilde \epsilon_i|^2 \frac{t^2}{(s-m_i^2)(m_i^2-u)}	\, ,	\label{eq:ComptonAmplitude}
\end{equation}
\begin{equation}
	\sum |\M|^2_{(b)} = 16\pi \hskip1pt A_\psi \, |\tilde \epsilon_i|^2 \frac{s^2}{(m_i^2-t)(m_i^2-u)} \, ,		\label{eq:annihilationAmplitude}
\end{equation}
with $\tilde{\epsilon}_i \equiv 2 m_i/\Lambda_i$, the Mandelstam variables~$s$, $t$ and~$u$, and
\begin{equation}
	A_\psi \equiv \left\{\begin{array}{ll} \displaystyle \alpha & \qquad \psi = \mathrm{lepton}, \\[8pt] 4 \alpha_s & \qquad \psi = \mathrm{quark}. \end{array} \right. 	
\end{equation}
We will neglect the weak temperature dependence of the running fine-structure constant~$\alpha$ and approximate it by its low-energy value of $\alpha \approx 1/137$. On the other hand, the running strong coupling constant~$\alpha_s(T)$ significantly depends on temperature. We include this temperature dependence by employing the five-loop corrections of the QCD~beta function implemented in \texttt{RunDec}~\cite{Herren:2017osy} for all temperatures with $\alpha_s < 1$. In the following, we will (conservatively) stop our calculation at a temperature of $T = \SI{1}{GeV}$ when $\alpha_s \approx 0.5$ to avoid the strongly-coupled regime. We refer to Appendix~\ref{app:rateComparisonsUncertainties} for additional details, a discussion of the implications of this choice and a less conservative calculation.\medskip

In general, the production rate for the relevant two-to-two processes is
\begin{align}
	\Gamma_\phi = \frac{1}{n^\mathrm{eq}_\phi} \prod_{i=1}^{4} \int\! \frac{\d^3 p_i}{(2\pi)^3 2 E_i}&\, f_1(p_1)\, f_2(p_2) \left[1 \pm f_3(p_3)\right] \left[1 \pm f_4(p_4)\right]	\label{eq:productionRate}	\\
						& \times (2\pi)^3 \delta^{(3)} (\p_1 + \p_2 - \p_3 - \p_4) \, (2\pi) \delta(E_1 + E_2 - E_3 - E_4)\, \sum |\mathcal{M}|^2\, , \nonumber
\end{align}
where~$n_\phi^\mathrm{eq}(T) = \zeta(3)\, T^3 / \pi^2$ is the equilibrium number density of a relativistic scalar at temperature~$T$, the momenta and energies of the incoming (outgoing) particles are denoted by~$\p_i$ and~$E_i$ with $i=1,2$~($3,4$), the Bose-Einstein and Fermi-Dirac distribution functions for bosons and fermions are
\begin{equation}
	f^b(p) = \frac{1}{\ee^{E(p)/T} - 1}\, , \qquad \, f^f(p) = \frac{1}{\ee^{E(p)/T} + 1}\, ,	\label{eq:quantumDistributions}
\end{equation}
and `$\pm$' indicates either Bose enhancement~(`$+$') or Pauli blocking~(`$-$') of the outgoing bosons and fermions, respectively. The total scattering amplitude is given by $\sum |\M|^2 = 2 \sum |\M|^2_{(a)} + \sum |\M|^2_{(b)}$ to account for fermions and antifermions in the Compton-like process. We show in Appendix~\ref{app:computationalDetails} that the rate~\eqref{eq:productionRate} can be rewritten as the following five-dimensional integral:
\begin{align}
	\Gamma_\phi = \frac{1}{n^\mathrm{eq}_\phi} \int_{E_\mathrm{min}}^\infty\! \d E \int_0^{p_\mathrm{max}}\! \d p \int_{p_1^\mathrm{min}}^{p_1^\mathrm{max}}\! \d p_1 \int_{p_3^\mathrm{min}}^{p_3^\mathrm{max}}\! \d p_3 &\, \frac{p_1\hskip1pt p_3}{512\pi^6 E_1 E_3} f_1(p_1) f_2(p_2)	\label{eq:productionRate2}	\\
						& \times \left[1 \pm f_3(p_3)\right] \left[1 \pm f_4(p_4)\right]\, \int_0^{2\pi}\! \d\phi\,\sum|\mathcal{M}|^2\, ,	\nonumber
\end{align}
where~$E$ and~$p$ are the total energy and momentum, $p_1$ and~$p_3$ are one of the incoming and outgoing momenta each, and $\phi$~is the polar angle difference between these two momenta in the plane orthogonal to~$\p$. We implicitly impose energy-momentum conservation to fix~$\p_{i+1} = \p - \p_i$ for $i=1,3$ and provide the integration limits in Appendix~\ref{app:computationalDetails}.

\begin{figure}
	\centering
	\includegraphics{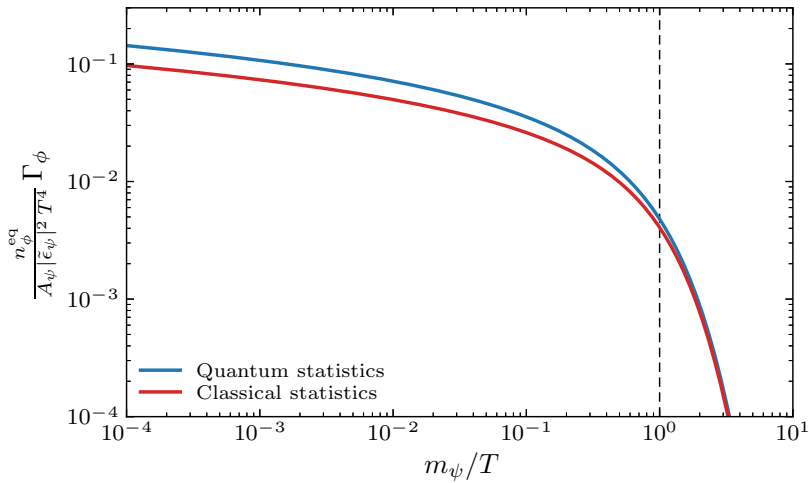}\vspace{-0.05in}
	\caption{Dimensionless rescaling of the interaction rate~$\Gamma_\phi$ as a function of~$m_\psi/T$. We compare the results of our full calculation using the Bose-Einstein and Fermi-Dirac distribution functions~(`quantum statistics') with the approximate result of employing the Boltzmann distribution without Bose enhancement and Pauli blocking~(`classical statistics'). The vertical dashed line indicates $T=m$ which is approximately the temperature where decoupling occurs for moderate coupling strengths.}
	\label{fig:dimensionlessInteractionRate}
\end{figure}

For the specific amplitudes of Compton-like scattering~\eqref{eq:ComptonAmplitude} and fermion annihilation~\eqref{eq:annihilationAmplitude} that are of interest in this work, the integral over the angle~$\phi$ can be conducted analytically. This means that we are left with a four-dimensional integral which we evaluate numerically using multi-dimensional adaptive quadrature. To facilitate its numerical calculation, it is useful to consider the rescaling~$n_\phi^\mathrm{eq} A_\psi^{-1} |\tilde{\epsilon}_\psi|^{-2} T^{-4}\, \Gamma_\phi$ as a function of~$m_\psi/T$ since it is dimensionless and independent of both the axion-fermion coupling and the SM~fermion masses. While we take these masses~$m_\psi$ to be non-zero, we assume massless axions, $m_\phi = 0$, which is a good approximation for large parts of parameter space relevant for measurements of the relativistic energy density as parameterized by~$\Neff$. The result is shown in Fig.~\ref{fig:dimensionlessInteractionRate}%
for the full quantum distribution functions of~\eqref{eq:quantumDistributions} with Bose enhancement and Pauli blocking, and the commonly employed classical approximation of Boltzmann statistics, $f^b(p) = f^f(p) = \exp\!{\left\{-E(p)/T\right\}}$, without Bose enhancement or Pauli blocking. As expected, the curves agree in the Boltzmann-suppressed regime of low temperatures, $T \lesssim m_\psi$ within~20\%, but differ at large temperatures $T \gg m_\psi$ (e.g.\ about~50\% for $T = \num{e3}\,m_\psi$). Since the difference in the range $m_\psi/T \in [1,10]$, which is most relevant for equilibrium physics, roughly varies between~15\% and~20\%, we expect shifts of less than about~10\% in our final predictions for the contribution to~$\Neff$, except in their tails at small couplings where the differences may be considerably larger.

\subsection{Computation of Constraints}
\label{sec:constraints}

Having calculated the production rate as a function of temperature~$T$, we now compute the number density~$n_\phi$ of axions and other~pNGBs, and the associated contribution to~$\Delta\Neff$ as a function of the coupling~$\Lambda_i$. This subsequently allows us to provide bounds on axion-fermion interactions from current and future cosmological measurements of~$\Delta\Neff$.\vspace{4pt}

Instead of assuming instantaneous annihilation of the SM~fermions at $T = m_\psi$ to estimate the relevant axion abundance, we solve the Boltzmann equation for the axion number density~$n_\phi$,\vspace{-3pt}
\begin{equation}
	\frac{\d n_\phi}{\d t} + 3 H(T)\, n_\phi = \Gamma_\phi(T)\, \big(n_\phi^\mathrm{eq}(T)-n_\phi\big)\, ,	\label{eq:BoltzmannEquation}
\end{equation}
with the Hubble parameter~$H(t)$ during radiation domination. While our calculation of the production rate~$\Gamma_\phi(T)$ is general, including Bose enhancement and Pauli blocking, this equation assumes that this quantity is determined by the rate in equilibrium and is therefore independent of~$n_\phi$. We only expect a minor impact of these assumptions for coupling strengths~$\Lambda_i$ for which the axion reaches equilibrium at temperatures $T \gtrsim m_\psi$, i.e.~when the expected number density of axions is near its equilibrium value. On the other hand, if the axion-fermion coupling is so small that the axion is never close to reaching equilibrium, our calculation of~$\Gamma_\phi(T)$ leads to a slight overestimate of this rate because it includes the equilibrium Bose enhancement from using~$n_\mathrm{eq}(T)$ in the final state.\footnote{We could have alternatively solved the Boltzmann equation for the distribution function~$f_\phi(p,T)$, which is an integro-differential equation, instead of the respective equation~\eqref{eq:BoltzmannEquation} for the number density~$n_\phi(T)$ to capture this effect.} In consequence, our calculations may overestimate the contribution to the radiation density as a function of the coupling strength,~$\Delta\Neff(\Lambda_i)$, by about~30\% around $\Delta\Neff=0.02$ or, alternatively, the bound on the interaction strength given a $\Delta\Neff$~measurement,~$\Lambda_i(\Delta\Neff)$, by roughly~10\% (with the latter being the quantity that we are more interested~in).

We follow the common procedure (see e.g.~\cite{DEramo:2018vss, Ferreira:2018vjj, Arias-Aragon:2020shv}) of numerically solving this differential equation after changing variables to the dimensionless inverse temperature $x = m_\psi/T$ and the dimensionless comoving number density~$Y_\phi = n_\phi/s$, where $s=2\pi^2g_{*s}\,T^3\!/45$ is the entropy density. We adopt the effective number of degrees of freedom in entropy~$g_{*s}(T)$ as numerically computed by~\cite{Saikawa:2018rcs}, which is based on the lattice QCD~calculation of~\cite{Borsanyi:2016ksw} in the non-perturbative regime,\footnote{If we used the results of~\cite{Borsanyi:2016ksw} over the entire temperature range, our predictions for~$\Delta\Neff$ would be within~$\lesssim 10\%$ of the presented results.} and an initial condition with no axions, $Y_{\phi,0} \equiv Y_{\phi, t=0} = 0$. Having obtained the final value for the comoving number density,~$Y_{\phi,\infty}$, as a function of the SM~fermion~$\psi_i$ and the axion-fermion interaction strength~$\Lambda_i$, we can convert it to a prediction for the contribution to the radiation density according to $\Delta\Neff \approx 74.84\, Y_{\phi,\infty}^{4/3}$. (We refer to Appendix~\ref{app:computationalDetails} for additional details.)\vspace{4pt}

\begin{figure}
	\centering
	\includegraphics{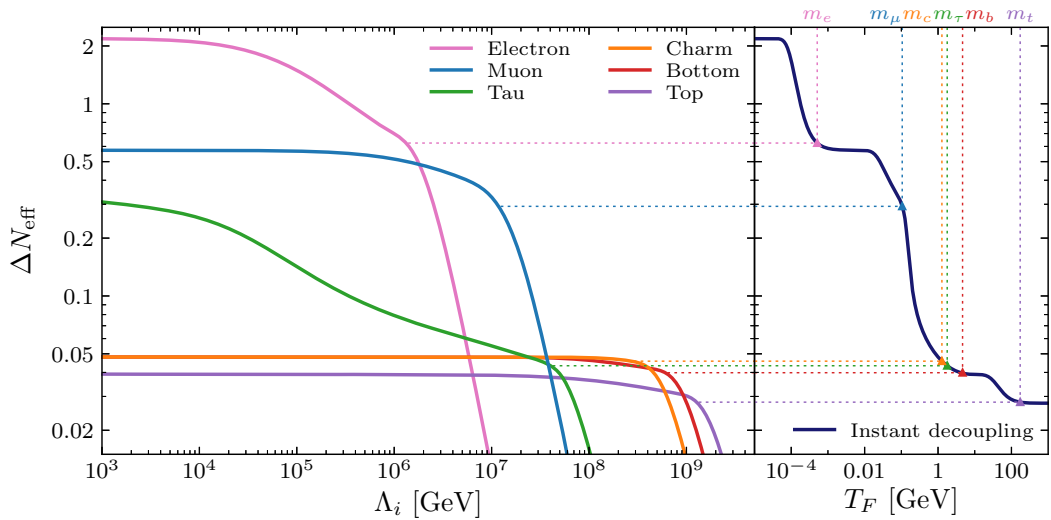}
	\caption{\textit{Left:} Contribution to the radiation density as parameterized by~$\Delta\Neff$ as a function of the axion-fermion coupling strength~$\Lambda_i$ for different SM~fermions~$\psi_i$. The displayed values for the bottom and charm couplings are conservative and may be (significantly) larger, with details of these uncertainties being discussed in Appendix~\ref{app:rateComparisonsUncertainties}. \textit{Right:} Contribution to~$\Delta\Neff$ for a single thermalized (equilibrium) degree of freedom which decoupled from the Standard Model at a temperature~$T_F$ (i.e.~the same as Fig.~\ref{fig:DeltaNeff_freezeout}). The horizontal lines between the panels indicate the contribution to~$\Delta\Neff$ expected for $T_F = m_\psi$ on the right with the appropriate value of~$\Lambda_i$ on the~left. For larger value of~$\Lambda_i$, the particle fails to reach equilibrium and, therefore, the abundance decreases rapidly.}
	\label{fig:DeltaNeff_lambda_extended}
\end{figure}

The results of this calculation are presented in Fig.~\ref{fig:DeltaNeff_lambda_extended}, %
which shows the contribution to~$\Delta\Neff$ for each of the axion couplings to SM~fermions.\footnote{The differences between our results obtained using full quantum statistics and calculations based on approximate classical statistics is about~5\% in the plateaus, but increases when the predicted values of~$\Delta\Neff$ drop for larger values of~$\Lambda_i$, reaching or even exceeding~20\% at $\Delta\Neff = 0.02$. This is because the axions decouple at higher temperatures where the difference between the classical and quantum production rate becomes more pronounced, cf.~Fig.~\ref{fig:dimensionlessInteractionRate}.} Given the current constraint from Planck, \mbox{$\Delta\Neff < 0.30$}~(95\%)~\cite{Planck:2018vyg}, we can constrain the axion coupling to electrons, muons and tau~leptons:\footnote{Apart from bounds on~$\Delta\Neff$ from the~CMB, we can also employ its BBN~constraints (see e.g.~\cite{Cyburt:2015mya, Pitrou:2018cgg, Fields:2019pfx}) which lead to
\begin{align}
	\Lambda_e 		&> \SI{1.8e6}{GeV} = \SI{e6.2}{GeV}\, ,	\label{eq:bound_electron_BBN}	\\
	\Lambda_\mu		&> \SI{1.4e6}{GeV} = \SI{e6.1}{GeV}\, ,	\label{eq:bound_muon_BBN}
\end{align}
where we conservatively assumed $\Delta\Neff < 0.5$ following~\cite{Ghosh:2020vti}. While the electron bound is similar due to the functional dependence of~$\Delta\Neff$ on~$\Lambda_e$ in the Boltzmann-suppressed regime, the muon constraint is weaker by one order of magnitude. On the other hand, these BBN~bounds are not limited to sub-\si{eV} axions, but extend to masses~$m_\phi \lesssim \SI{1}{MeV}$ (cf.~\cite{Ghosh:2020vti}).}
\begin{align}
	\Lambda_e		&> \SI{2.5e6}{GeV} = \SI{e6.4}{GeV}\, ,	\label{eq:bound_electron_Planck}\\
	\Lambda_\mu		&> \SI{1.1e7}{GeV} = \SI{e7.1}{GeV}\, ,	\label{eq:bound_muon_Planck}	\\
	\Lambda_\tau	&> \SI{1.7e3}{GeV} = \SI{e3.2}{GeV}\, .	\label{eq:bound_tau_Planck}
\end{align}
Upcoming (near-term) CMB~experiments will continue to improve the measurement of~$\Neff$ which will also increase the sensitivity to these interactions. Given that the Simons Observatory~\cite{SimonsObservatory:2018koc} and \mbox{CMB-S4}~\cite{Abazajian:2019eic} are forecasted to reach $\Delta\Neff < 0.14$ and~$0.060$ at 95\%~c.l., respectively, we project that~SO can exclude
\begin{align}
	\Lambda_e		&> \SI{3.7e6}{GeV} = \SI{e6.6}{GeV}\, ,	\label{eq:bound_electron_SO}	\\
	\Lambda_\mu		&> \SI{2.2e7}{GeV} = \SI{e7.3}{GeV}\, ,	\label{eq:bound_muon_SO}		\\
	\Lambda_\tau	&> \SI{1.0e5}{GeV} = \SI{e5.0}{GeV}\, , \label{eq:bound_tau_SO}
\end{align}
while \mbox{CMB-S4} will just fall short of the heavy quark targets,\footnote{We note the relatively large uncertainty in our predictions for the coupling to the charm and bottom~quarks due to the strong-coupling regime of the QCD~phase transition (see Appendix~\ref{app:rateComparisonsUncertainties} for a more detailed discussion). Dedicated QCD~lattice calculations may reveal that~\mbox{CMB-S4} and potentially even~SO are sensitive to these interactions. An alternative approach is to match across the QCD~phase transition using calculations for~$\Delta \Neff$ before and after, similar to the calculation of~\cite{DEramo:2021psx, DEramo:2021lgb} for the QCD~axion.} but will constrain
\begin{align}
	\Lambda_e		&> \SI{5.4e6}{GeV} = \SI{e6.7}{GeV}\, ,	\label{eq:bound_electron_S4}	\\
	\Lambda_\mu		&> \SI{3.3e7}{GeV} = \SI{e7.5}{GeV}\, ,	\label{eq:bound_muon_S4}		\\
	\Lambda_\tau	&> \SI{5.5e6}{GeV} = \SI{e6.7}{GeV}\, .	\label{eq:bound_tau_S4}
\end{align}
We in particular note the improvement in~$\Lambda_\tau$ by nearly two orders of magnitude when going from Planck to~SO and from~SO to~\mbox{CMB-S4}, respectively. While the physical origins of these bounds will be discussed in~\textsection\ref{sec:physics}, we clearly see that there is a highly nonlinear relationship between improvements in the measurement of~$\Neff$ and the parameters of models that produce $\Delta\Neff > 0$.\medskip

These bounds hold for axions with masses $m \lesssim \SI{1}{eV}$. At higher masses, the axions behave like matter (and not free-streaming radiation) during the recombination era and may also be constrained from their contribution to the effective mass of neutrinos. As we increase the axion mass, we expect that the constraints on these couplings will become more stringent due to their impact on structure formation and, eventually, over-closure of the universe. However, as we increase the mass, it is increasingly possible that the axions decay prior to recombination through a coupling to photons or neutrinos. We will therefore leave the discussion of larger masses to future work~(see also~\cite{Millea:2015qra, DePorzio:2020wcz, Xu:2021rwg}).

\subsection{Physics of Constraints}
\label{sec:physics}

In the previous subsection, we calculated~$\Delta\Neff$ for axions and other~pNGBs coupled to individual SM~fermions. The shapes of the curves displayed in Fig.~\ref{fig:DeltaNeff_lambda_extended} vary significantly depending on the fermion. In the following, we provide qualitative explanations and describe the physics underlying this functional dependence of~$\Delta\Neff(\Lambda_i)$. In addition, we might hope to understand the approximate size of the constraints on the interaction strengths from dimensional analysis.\medskip

As described in detail in the previous subsection, cosmological production of axions is described by the Boltzmann equation~\eqref{eq:BoltzmannEquation}. While the total production of axions can be determined exactly using this equation, the origin of the constraint follows from the qualitative requirement that the production becomes efficient at some temperature. Specifically, we will produce a large number density of axions if the production rate exceeds the rate of dilution due to the expansion of the universe, namely
\begin{equation}
	\Gamma(T) > H(T) = \sqrt{\frac{\pi^2}{90}g_*(T)}\,\frac{T^2}{\Mpl} \, ,	\label{eq:condition}
\end{equation}
for some temperature~$T$ achieved in the early universe, with the reduced Planck mass~$\Mpl$. When this condition is met, the axion will thermalize, i.e.\ the number density of axions will approach the number density of photons at that temperature which can therefore yield a potentially detectable contribution to~$\Neff$.

\begin{figure}
	\centering
	\includegraphics{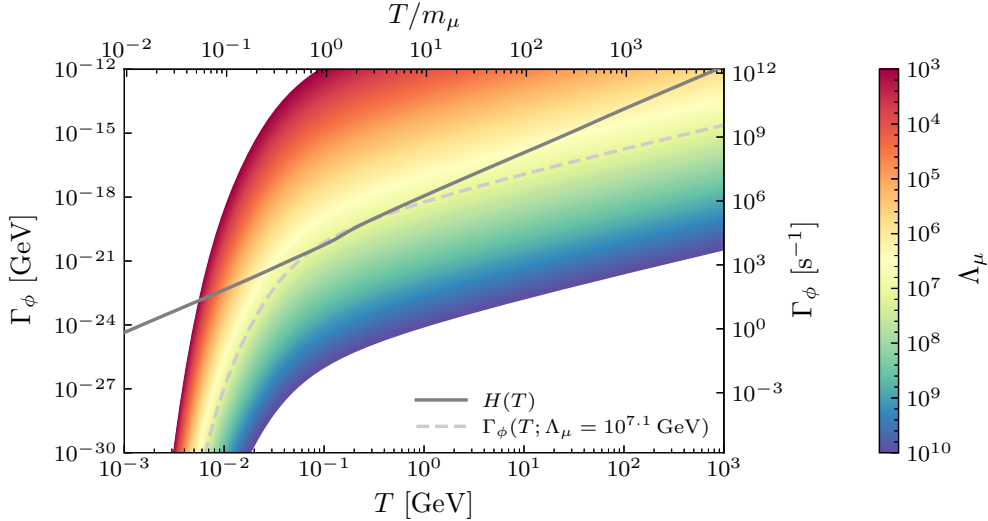}
	\caption{Production rate of axions and other~pNGBs,~$\Gamma_\phi$, as a function of temperature~$T$ for different values of the muon coupling~$\Lambda_\mu$. The solid gray line shows the Hubble rate~$H(T)$, i.e.\ $\Gamma_\phi(T) > H(T)$ indicates efficient cosmological production of axions at that temperature. The dashed line shows the production rate for the current bound on the interaction strength from $\Neff$~measurements of~Planck. For small values of~$\Lambda_\mu$, we see efficient production over many decades of $T > m_\mu$, but the axion usually decouples by $T \lesssim m_\mu/10$ because the production becomes exponentially suppressed. As~$\Lambda_\mu$ increases, efficient production is increasingly possible only around $T \approx m_\mu$. When $\Lambda_\mu \gg \SI{e7}{GeV}$, there is no temperature where axions are efficiently produced from this interaction with muons. While the most natural units of the production rate and temperature for cosmological axion production are~\si{GeV}~(left) and the dimensionless ratio~$T/m_\mu$~(top), production rates in~\si{\per\second}~(right) and temperatures in~\si{GeV}~(bottom) are useful for comparison with astrophysical axion production.}
	\label{fig:productionRate}
\end{figure}

At any temperature above the mass of the fermion, we can always make the coupling sufficiently large to meet our condition in~\eqref{eq:condition} which implies that the axions will thermalize. This is sufficient to ensure that the axions remain in thermal equilibrium with the rest of the Standard Model as the universe cools to lower temperatures (but that are still larger than the mass, $T > m_i$). However, due to the Boltzmann suppression of the fermion when $T \ll m_i$, the axions always decouple from the~SM bath at sufficiently low temperatures, regardless of the coupling strength. Figure~\ref{fig:productionRate}%
compares the Hubble rate~$H(T)$ to the production rate~$\Gamma(T)$ for a range of interaction strengths to muons and illustrates both of these features. We notice that all production rates become much smaller than the Hubble rate, $\Gamma_\phi(T) \ll H(T)$, by $T \approx m_\mu/10$. As a consequence, we should expect that the left-most part of Fig.~\ref{fig:DeltaNeff_lambda_extended} should asymptote to~$\Delta\Neff$ as determined by a decoupling at $T_F \approx m_i/10$ based on Fig.~\ref{fig:DeltaNeff_freezeout}. This implies that the asymptotic contributions to~$\Delta\Neff$ for large couplings to different fermions should be ordered inversely proportional to their mass, i.e.~smaller masses~$m_i$ correspond to larger~$\Delta\Neff$ because $T_F \approx m_i/10$ is smaller. We see that this ordering is true for all the heavy fermions except for the ordering of the charm and tau~lepton. This is a result of cutting off the axion production rate during the QCD~phase transition when coupling to the charm because the strong coupling constant is no longer perturbative. This uncertainty is discussed in Appendix~\ref{app:rateComparisonsUncertainties} and \mbox{implies that the charm and bottom curves are likely underestimated at larger couplings.}

As we decrease the coupling, or increase the scale~$\Lambda_i$, we see that the shape of the curves in Fig.~\ref{fig:DeltaNeff_lambda_extended} depends significantly on the specific fermion. Increasing the coupling changes the precise temperature at which the axion decouples and, therefore, the part of the $g_*(T)$~curve responsible for diluting the number of axions. Specifically, as the coupling decreases, the temperatures where the axion production is significant becomes increasingly restricted to $T \approx m_i$. As a consequence, the decoupling temperature effectively increases from $T_F \approx m_i/10$ to $T_F \approx m_i$ as we move from stronger to weaker coupling (smaller to larger~$\Lambda_i$). This implies that the contribution to~$\Delta\Neff$ from coupling to a given fermion is sensitive to the $g_*(T)$~curve in the vicinity of its mass~$m_i$. We note that the contribution of the fermion itself is included in this change to~$g_*(T)$ which means that there are more fermion-antifermion pairs present in the thermal bath when the axions decouple at higher temperatures. In all cases, we see a knee in the shape of the curve that is in good agreement with the equilibrium result for $T_F = m_i$, which is the approximate decoupling temperature when the axions just barely reach equilibrium.

Finally, as we decrease the coupling further (again, equivalent to increasing~$\Lambda_i$), the production rate will eventually not reach~$H(T)$ for any temperature~$T$. Without coming into equilibrium, the number of axions is no longer tied to the number of photons and we see that the contribution to~$\Delta\Neff$ falls rapidly. Since the effective coupling is~$\tilde{\epsilon}_i = 2m_i/\Lambda_i$, this happens at smaller values of~$\Lambda_i$ for lighter fermions. As a consequence, the exponential falloff of the $\Delta\Neff$~curves in Fig.~\ref{fig:DeltaNeff_lambda_extended} occurs in the order of increasing mass, i.e.~lighter fermions lead to negligible~$\Delta\Neff$ at smaller~$\Lambda_i$. The fact that the $\Delta\Neff$~curves cross is another manifestation of the same physics.

We can gain further insights by comparing the shape of the curves relating~$\Delta\Neff$ and the axion-fermion coupling strengths (left panel of Fig.~\ref{fig:DeltaNeff_lambda_extended}) to the standard $T_F$-$\Delta\Neff$ curve of Fig.~\ref{fig:DeltaNeff_freezeout}. This comparison is provided by the right panel of Fig.~\ref{fig:DeltaNeff_lambda_extended}. Taking $T_F = m_i$, which is denoted by the triangles in the right panel, we see that the contributions to~$\Delta\Neff$ are near, but slightly below the asymptotic values at strong couplings for most fermions. While this is in line with our expectations, it does not fully address how to translate the approximate inequality in~\eqref{eq:condition} into a map between~$\Lambda_i$ and~$\Delta\Neff$. This is particularly noticeable for the coupling to muons, where the $\Delta\Neff$~curve begins to drop around $\Lambda_\mu = \SI{e6}{GeV}$, but crosses~$\Delta\Neff(T_F = m_\mu)$ only around $\Lambda_\mu = \SI{e7}{GeV}$. More generally, the limitation of our qualitative estimates is that they do not entirely explain the shapes of the $\Delta\Neff$~curves.\medskip

\begin{figure}
	\centering
	\includegraphics{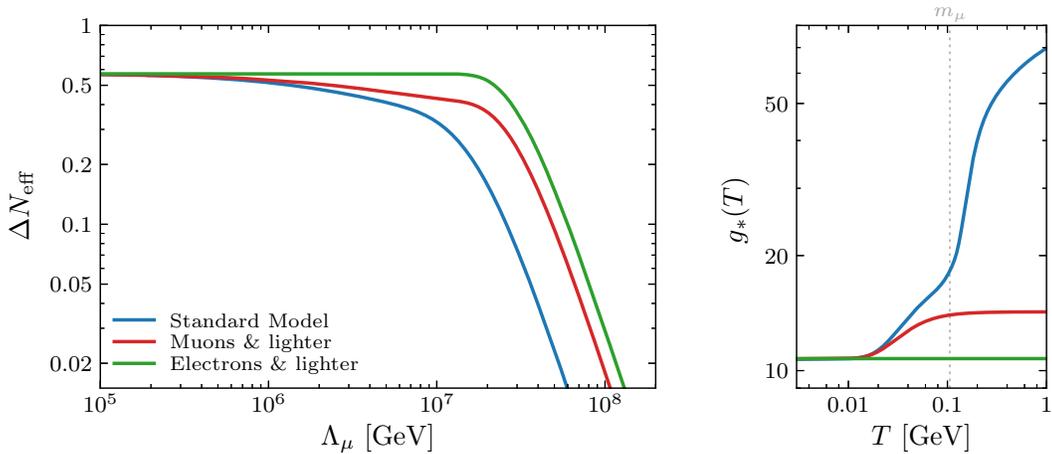}
	\caption{\textit{Left}:~Breakdown of the contribution to~$\Delta\Neff$ from the coupling to muons. The full Standard Model result is shown in blue, the red curve displays the predictions for a universe without the SM~fermions that are heavier than the muon, i.e.\ a universe with photons, neutrinos, electrons and muons, and the green curve additionally removes the contribution to~$g_*(T)$ from the muon. We see that the large change in~$g_*(T)$ in the vicinity of $T = m_\mu$ due to the QCD~phase transition has a significant impact on the resulting contribution to~$\Delta\Neff$. \textit{Right}:~The~$g_*(T)$~curves underlying the calculation of~$\Delta\Neff(\Lambda_\mu)$ displayed in the left panel, matched by color.}
	\label{fig:DeltaNeff_lambda_muons}
\end{figure}

The complication in relating the scales~$\Lambda_i$ to values of~$\Delta\Neff$ is that these curves are really a combination of the axion production rate from the fermion involved in the coupling and the effective number of relativistic degrees of freedom from all the particles in the Standard Model,~$g_*(T)$. We can separate these effects by removing the dependence on~$g_*(T)$, as is shown in Fig.~\ref{fig:DeltaNeff_lambda_muons}. %
The green curve shows the contribution to~$\Delta\Neff$ if~$g_*(T)$ was a constant over the relevant range of temperatures so that~$\Delta\Neff$ would be a constant for any coupling reaching equilibrium. We see in Fig.~\ref{fig:productionRate} that $\Gamma(T) < H(T)$ for all temperatures~$T$ if $\Lambda_\mu > \SI{1.5e7}{GeV} \approx \SI{e7.2}{GeV}$, which is in good agreement with the value of~$\Lambda_\mu$ at which the curve bends, indicating that the axions are never coming into equilibrium.

In contrast, even when we add the contribution to~$g_*(T)$ from the muon (displayed by the red curve in Fig.~\ref{fig:DeltaNeff_lambda_muons}), which changes between $T \gg m_\mu$ and $T \ll m_\mu$, the contribution to~$\Delta\Neff$ begins to decrease around $\Lambda_\mu \approx \SI{e6}{GeV}$ when the axion is still reaching equilibrium. We can understand this in terms of our simple estimate as follows: at large coupling strengths, $T_F \approx m_\mu / 10$ means the muons have already mostly annhilated at decoupling and do not contribute to~$g_*(T_F)$. As we decrease the coupling, $T_F \to m_\mu$ so that the axions are diluted somewhat by the muon annihilation even though the axions come into thermal equilibrium.

Finally, when all the degrees of freedom of the Standard Model are included in~$g_*(T)$ (blue curve in Fig.~\ref{fig:DeltaNeff_lambda_muons}), we see a considerably larger suppression even when $\Lambda_\mu > \SI{e7}{GeV}$ because the muon mass lies on the boundary of the QCD~phase transition during which the number of degrees of freedom is changing rapidly with temperature. This dependence on~$g_*(T)$ depends significantly on the fermion mass and leads to the variety of shapes seen in Fig.~\ref{fig:DeltaNeff_lambda_extended}. This is best illustrated by the coupling to the tau~lepton, which has a mass that is close to the QCD~phase transition which means that the $\Delta\Neff$~curve effectively transitions from decoupling after the QCD~phase transition to decoupling before the QCD~phase transition as we increase~$\Lambda_\tau$.

\section{Comparison with Other Probes}
\label{sec:comparison}

Cosmological constraints on axions and other~pNGBs are particularly compelling as they are both easy to calculate and robust to much of the details of the model. This is largely due to thermal equilibrium which tells us the number of axions produced at a given temperature for any sufficiently large coupling. Yet, there are a wide variety of other probes of axions, both terrestrial and astrophysical, that have different strengths and weaknesses compared to cosmological probes. In this section, we will compare our results, especially our constraint on~$\Lambda_\mu$ from Planck, to other probes of the same couplings. Of particular interest will be astrophysical constraints, such as from cooling of supernova~SN~1987A, which are also the result of thermal axion production.

\subsection{Stellar Cooling and SN~1987A}
\label{sec:stellarCooling}

Astrophysical constraints on axion couplings~\cite{Raffelt:1996wa, Raffelt:2006cw, Raffelt:2012kt} offer a useful foil for cosmological bounds. Stars also provide a controlled high-temperature environment in which the thermal production of a new light particle would be detectable. In this sense, astrophysical constraints are probing essentially the same physics as constraints from bounds on~$\Neff$. It is therefore instructive to understand where differences arise and what the relative strengths of each probe are.\medskip

At a qualitative level, both probes are sensitive to large changes in the number of axions. In the case of astrophysical environments, the production of these particles is governed by
\begin{equation}
	\frac{\d n_\phi}{\d t} = \Gamma_\phi^{(\star)}(T_\star) \left(n_\phi^\mathrm{eq}(T_\star) - n_\phi\right) ,
\end{equation}
where~$T_\star$ is a temperature that is (mostly) fixed by the specific probe. This equation should be compared to the cosmological Boltzmann equation~\eqref{eq:BoltzmannEquation}. In general, the cosmological interaction rate~$\Gamma_\phi(T)$ and the astrophysical rate~$\Gamma_\phi^{(\star)}(T)$ are related, but they can differ even at the same temperature due to the large chemical potentials present in astrophysical environments. This is particularly important for protons, neutrons and electrons. In contrast, muons (and the other heavier particles) are unstable and their abundance is primarily due to thermal production. As a result, we can treat $\Gamma_\phi(T) \approx \Gamma_\phi^{(\star)}(T)$ for our purposes.

At very weak coupling, the axions will escape the star after production, thus providing a new mechanism for energy to leave the system. If the number of axions produced in the timescale of observation~$t_\mathrm{obs}$ is comparable to the number of photons,
\begin{equation}
	\Gamma_\phi^{(\star)}(T_\star)\, n_\phi^\mathrm{eq}(T_\star)\, t_\mathrm{obs} \gtrsim n_\gamma(T_\star)\, ,
\end{equation}
then the energy loss due to axions is significant and would lead to detectable changes in the dynamics of the astrophysical system.\medskip

\begin{figure}
	\centering
	\includegraphics{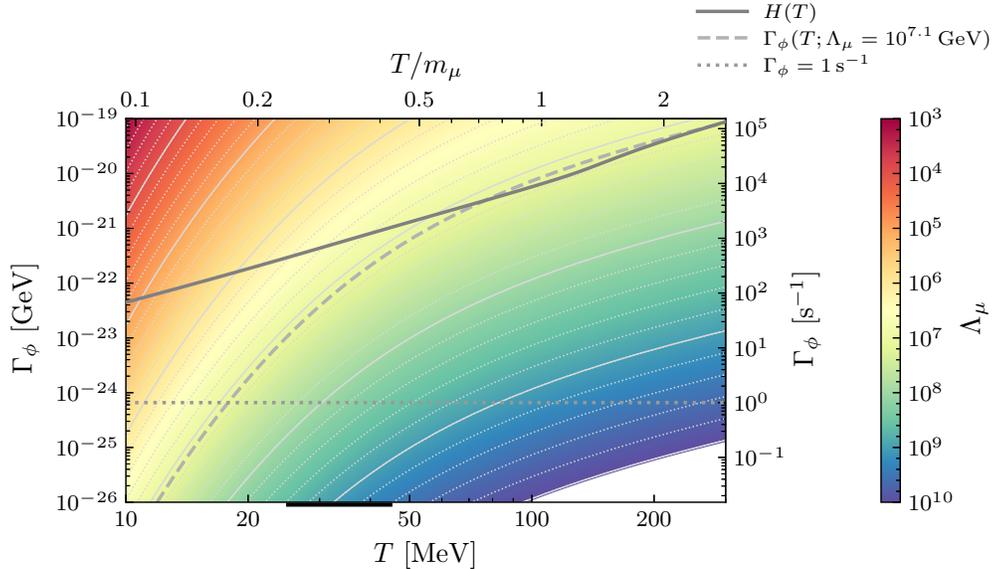}
	\caption{Production rate of axions and other~pNGBs,~$\Gamma_\phi$, as a function of temperature~$T$ for different choices of the coupling to muons,~$\Lambda_\mu$. As in Fig.~\ref{fig:productionRate}, which presents a larger range of temperatures, the solid gray line shows the Hubble rate~$H(T)$ to indicate the region $\Gamma_\phi(T) > H(T)$ of efficient cosmological production and the dashed gray line indicates the interaction rate for the current Planck~$\Neff$~bound on the axion-muon coupling~$\Lambda_\mu$. This figure allows us to easily compare our cosmological bounds to astrophysical constraints which arise for an order one change in the number of photons during the observable timescale, namely $\Gamma_\phi^{(\star)}(T)\, t_\mathrm{obs} > 1$. As~SN~1987A is associated with $t_\mathrm{obs} \approx \SI{1}{\second}$, the inferred constraint arises approximately from the intersection of $\Gamma_\phi \approx \Gamma_\phi^{(\star)} = \SI{1}{\per\second}$ with the temperature in the core of the supernova. As an estimate of this temperature, the black horizontal line indicates the range of mass-weighted, radially-averaged temperatures in spherically-symmetric, one-dimensional simulations of approximately~$\SIrange[open-bracket = [, close-bracket = ]]{25}{45}{MeV}$~\cite{Bollig:2020xdr}.}
	\label{fig:productionRate_zoom}
\end{figure}

We are particularly interested in the constraints from~SN~1987A because the high temperatures of the supernova can produce a large number of muons. The timescale relevant to axion cooling of the proto-neutron star is $t_\mathrm{obs} \approx \SI{1}{\second}$.\footnote{A common description of supernovae suggests the cooling of the proto-neutron star lasts approximately ten seconds. We will conservatively take the relevant timescale to be $t_\mathrm{obs} \approx \SI{1}{\second}$ because approximating supernovae as constant temperature systems will break down as $t_\mathrm{obs} \to \SI{10}{\second}$.} Of course, the temperature of the star depends both on the radius and time which therefore means that our estimate is necessarily approximate. Nevertheless, both the axion production and the total energy of the star are dominated by the hottest regions which therefore makes this local approximation a useful starting point. Furthermore, since $n_\phi^\mathrm{eq}(T_\star) \approx n_\gamma(T_\star)$ for light axions, forbidding significant cooling implies a constraint
\begin{equation}
	\Gamma_\phi^{(\star)}(T_\star) < t_\mathrm{obs}^{-1} = \SI{e-24.2}{GeV} = \SI{7e-25}{GeV}\, .
\end{equation}
For a system in equilibrium at temperature~$T$, we can easily determine this constraint from~Fig.~\ref{fig:productionRate_zoom}. %
We however notice that the production rate is very sensitive to the precise temperature. In particular, if we require that $\Gamma_\phi^{(\star)} = \SI{1}{\per\second}$, then our bound is in the range of $\Lambda_\mu > 10^{7.7-8.5}\,\si{GeV}$ for a temperature range of $\SIrange[open-bracket = [, close-bracket = ]]{25}{45}{MeV}$. This is, of course, a reflection of the Boltzmann suppression of muons inside the supernova and, therefore, explains our exponential sensitivity to the temperature. These results are in agreement with the bounds found by a more detailed analysis presented in~\cite{Bollig:2020xdr, Croon:2020lrf}. Using results from a simulation with a mass-weighted, radially-averaged core temperature of~$\SI{25}{MeV}$, they find $\Lambda_\mu > \SI{e7.5}{GeV}$. However, these bounds still depend sensitively on the specific simulation. Concretely, they find $\Lambda_\mu > \SI{e8.0}{GeV}$ in another simulation with mass-weighted, radially-averaged core temperature of~$\SI{45}{MeV}$. Importantly, the temperatures reached inside the supernova vary significantly both with position inside the core and between the simulations. As a result, we expect that the derived bounds will be exponentially sensitive to the details of the specific simulation used (see also~\cite{Caputo:2021rux} for more discussion).

From Figure~\ref{fig:productionRate_zoom}, we can also compare the sensitivity of the cosmological constraint to the bound from~SN~1987A. Comparing the production rate as a function of temperature~$T$ to the Hubble rate (shown by the solid gray line), we see that $\Lambda_\mu = \SI{e7.2}{GeV}$ is the smallest value for which the cosmological production never becomes efficient, $\Gamma_\phi(T) > H(T)$. This is our approximate cosmological bound that essentially reproduces the exact bound from current Planck measurements of~$\Delta\Neff$ of $\Lambda_\mu > \SI{e7.1}{GeV}$. To compare this constraint to~SN~1987A, we follow the dashed gray curve to $\Gamma_\phi = \SI{1}{\per\second}$ where it corresponds to the production in the supernova at $T = \SI{18}{MeV}$. In addition, we see that the temperature where $\Gamma_\phi = \SI{1}{\per\second}$ is nearly unchanged for the slightly smaller value of $\Lambda_\mu = \SI{e7}{GeV}$, but with the major difference that axions are efficiently produced in the early universe over many decades in temperature. In consequence, it is more useful to treat cosmological and supernova bounds of axions and other~pNGBs as complimentary rather than redundant. While the constraint $\Lambda_\mu> \SI{e7.5}{GeV}$ from~SN~1987A is somewhat stronger, the current cosmological limit of $\Lambda_\mu > \SI{e7.1}{GeV}$~(or the future \mbox{CMB-S4}~limit of $\Lambda_\mu> \SI{e7.5}{GeV}$) can be mostly understood from equilibrium physics and is therefore quite robust. In contrast, the temperatures and dynamics inside a supernova are complex and the origin of the constraints are more uncertain~(see also~\cite{Bar:2019ifz} for other potential limitations and uncertainties of the supernova-based constraints). For related reasons, the bounds in~\cite{Bollig:2020xdr, Croon:2020lrf} are theoretical constraints and not 95\%~c.l.~limits, in contrast to the cosmological constraints derived from $\Neff$~measurements.

\subsection{Experimental Limits}

We discussed the axion-muon coupling constraints from stellar cooling and~SN~1987A in detail in the previous section. In the following, we provide a more general overview of existing bounds on couplings of axions and other~pNGBs to matter and compare them to the limits that we derived in~\textsection\ref{sec:constraints}. We will keep our focus on the diagonal couplings and will not discuss the extensive list of existing bounds on off-diagonal couplings, \mbox{in particular from SM~particle decays~(cf.~e.g.~\cite{Feng:1997tn, Andreas:2010ms, Dolan:2014ska, Calibbi:2020jvd}).}

\subsubsection*{Axion-Lepton Couplings}

First, we will consider the couplings of axions to leptons. We focus on the axion-muon and axion-tau couplings since the~$\Delta\Neff$-based bounds are competitive in those cases, but will also briefly discuss the axion-electron interaction. Moreover, we are not considering model-dependent constraints that convert a bound on the coupling to electrons or photons to a limit on the other lepton couplings, e.g.\ by assuming these interactions to be universal or determined by a model such as DFSZ. Such bounds are generally stronger due to the tight bounds on~$\Lambda_e$, but we prefer to consider the various axion couplings to be independent. This results in conservative and model-independent estimates applicable to any pNGB-fermion interaction, as mentioned above.

\paragraph{Coupling to Electrons.}
The limit on the axion-electron interaction strength from white dwarf cooling is $\Lambda_e > \SI{1.2e10}{GeV}$~\cite{Hansen:2015lqa} (see also~\cite{Battich:2016htm, EDELWEISS:2018tde, Capozzi:2020cbu, Lucente:2021hbp, DiLuzio:2021ysg} for similar limits from stellar cooling and related discussions), which is considerably stronger than any limit that may be derived from upcoming cosmological~$\Delta\Neff$ measurements. The reheating temperature-dependent freeze-out constraint from excluding $\Delta\Neff = 0.027$ is $\Lambda_e \gtrsim \SI{6e7}{GeV} \sqrt{T_R/\SI{e10}{GeV}}$~\cite{Baumann:2016wac}, while the current freeze-in constraint is given by $\Lambda_e > \SI{2.5e6}{GeV}$ which will improve to~\SI{5.4e6}{GeV} with~\mbox{CMB-S4}, cf.~\eqref{eq:bound_electron_Planck} and~\eqref{eq:bound_electron_S4}.\footnote{We note that BBN-based $\Delta\Neff$~measurements have been able to essentially close a small window in parameter space for $m_\phi \in \SIrange[open-bracket = [, close-bracket = ]]{0.1}{1}{MeV}$~\cite{Ghosh:2020vti}.} Having said that, the cosmological $\Delta\Neff$-based bounds on~$\Lambda_e$ are less sensitive than astrophysical constraints to the physical environment where axions are produced. In addition, the environment in the early universe is quite different to the interiors of stars which means that cosmological probes can be an important complementary test~\cite{DeRocco:2020xdt}.

\paragraph{Coupling to Muons.}
The main constraints on potential interactions of sub-\si{MeV} axions and muons have been derived from the cooling rate of supernova~SN~1987A and from measurements of the anomalous magnetic moment of the muon,~$g_\mu-2$. While we discussed the former in detail in~\textsection\ref{sec:stellarCooling}~(cf.~\cite{Bollig:2020xdr, Croon:2020lrf}), the latter allows to put laboratory bounds on the coupling strength since~pNGBs contribute to~$g_\mu-2$ at the loop level. Following~\cite{Essig:2010gu} and using the current difference between the measured and theoretically-predicted value of~$a_\mu \equiv (g_\mu-2)/2$, $\Delta a_\mu = \num[separate-uncertainty=true]{251(59)e-11}$~\cite{Aoyama:2020ynm, Muong-2:2021ojo}, at the lower $5\sigma$~limit (since the pNGB-induced contribution to the anomalous moment is negative), $\Delta a_\mu \geq \num{-44e-11}$, we conservatively derive $\Lambda_\mu > \num{e2.6}$.\footnote{We note that the difference between the experimental value and the implied value from recent lattice calculations of the hadronic contribution to~$g_\mu-2$ is significantly smaller~\cite{Borsanyi:2020mff} and leads to a bound which is weaker by a factor of approximately two.}

\begin{figure}
	\centering
	\includegraphics{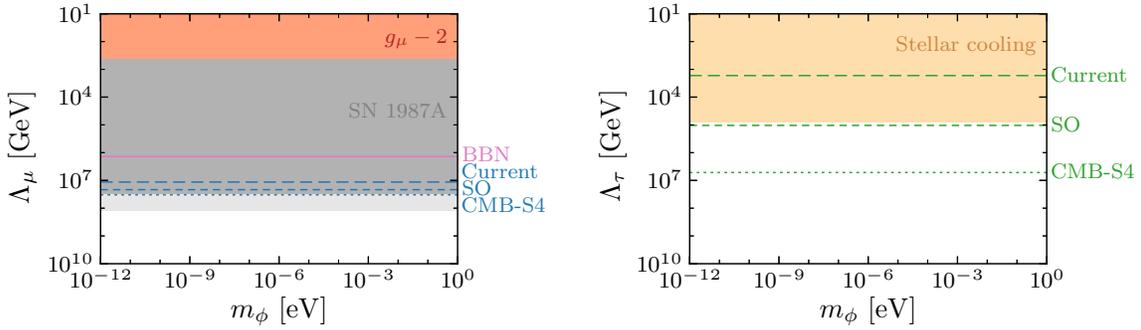}
	\caption{Comparison of existing and future constraints on the coupling between axions and muons~(\textit{left}) and tau~leptons~(\textit{right}), respectively, as a function of the axion mass~$m_\phi \leq \SI{1}{eV}$, which is the mass range relevant for CMB~(and large-scale structure)~measurements of~$\Delta\Neff$. \textit{Left}: The existing model-independent constraints for the coupling to muons come from measurements of the anomalous magnetic moment of the muon ($g_\mu-2$), which would receive axion contributions at the loop level, and from the observed cooling rate of~SN~1987A. For the latter, the dark (light) regions indicate the conservative (optimistic) bounds inferred by~\cite{Bollig:2020xdr, Croon:2020lrf} and, therefore, indicate the level of uncertainty in these constraints. The limits derived in this work from $\Delta\Neff$~measurements of~BBN (cf.~\cite{Ghosh:2020vti}), and current and future CMB~experiments are complementary in nature. \textit{Right:} The strongest, model-independent bound on the axion-tau interaction strength comes from the loop-induced coupling to electrons which is strongly constrained from stellar cooling of white dwarfs. The CMB-based limits will improve by two orders of magnitude each from Planck to the Simons Observatory and~\mbox{CMB-S4}.}
	\label{fig:axionLeptonConstraints}
\end{figure}

In the left panel of Fig.~\ref{fig:axionLeptonConstraints},%
we compare these existing bounds to the current and future bounds studied in this work, see~\eqref{eq:bound_muon_BBN}, \eqref{eq:bound_muon_Planck}, \eqref{eq:bound_muon_SO} and~\eqref{eq:bound_muon_S4}. For the supernova bound, we display a conservative and an optimistic estimate as derived in~\cite{Bollig:2020xdr, Croon:2020lrf} which differ in their model assumptions of the mass of the remnant neutron star in~SN~1987A, with additional uncertainties possible from uncertainties in the equation of state at supernuclear densities. We clearly see the complimentary nature of the displayed constraints, with current and future $\Delta\Neff$-based bounds exploring the same parameter space and an exclusion of $\Delta\Neff > 0.067$ corresponding to the conservative bound from~SN~1987A.

\paragraph{Coupling to Tau Leptons.}
In contrast to the interactions with electrons and muons, there do not appear to be model-independent, tree-level bounds on the diagonal coupling to tau~leptons. However, such a coupling would induce an interaction with electrons at the loop level, i.e.\ the bounds on~$\Lambda_e$ can generally be translated into constraints on the couplings to the other leptons~(and quarks)~\cite{Feng:1997tn}. The strong bounds on the electron-axion coupling and the large masses of the third generation of SM~fermions partly compensate the loop suppression which results in interesting constraints. Following~\cite{Feng:1997tn}, we infer $\Lambda_\tau \gtrsim \SI{8e4}{GeV}$ from the previously mentioned white dwarf cooling bound on~$\Lambda_e$ of~\cite{Hansen:2015lqa} if the axion contribution to the induced coupling is dominantly proportional to~$m_\tau^2$. We compare this limit to our freeze-in constraints in the right panel of Fig.~\ref{fig:axionLeptonConstraints}. While the current Planck bound is weaker than the loop-induced bound based on the $\Lambda_e$~limit, near-term CMB~experiments will strengthen this constraint, with the sensitivity of~SO corresponding to this stellar bound and \mbox{CMB-S4}~being projected to improve upon it by about two orders of magnitude.

\medskip
Overall, we notice that the cosmological constraints have the opposite strengths and weakness of the astrophysical bounds on axion couplings to leptons. Due to the maximum temperatures found in astrophysical settings, the implied sensitivities are much weaker for heavier leptons. In contrast, under plausible assumptions, our cosmological history reaches temperatures well above the masses of these leptons and, therefore, is also sensitive to the heavy leptons. In fact, cosmological observations can reach larger values of~$\Lambda_i$ for the heavier fermions because the effective coupling is proportional to their mass.

\subsubsection*{Axion-Quark Couplings}

We now turn to the interaction between axions and the heavy quarks. Since current and near-term cosmological experiments are not sensitive enough to constrain these couplings,\footnote{Note however our discussion in Appendix~\ref{app:rateComparisonsUncertainties} of the uncertainties in our calculation for the interactions with the charm and bottom~quarks which still leave the possibility for such constraints.} we will instead estimate the required sensitivity to~$\Delta\Neff$ to match the existing constraints. As for the lepton couplings, we will again focus on model-independent and diagonal couplings, but note that it is less clear in this case because the distinction between diagonal and off-diagonal constraints is only valid at leading order since quark flavors necessarily mix.

\paragraph{Coupling to Top Quarks.}
The best model-independent constraints on the diagonal axion-top coupling arise from the loop-induced constraint based on the $\Lambda_e$~limit, cf.~\cite{Feng:1997tn}. Assuming that this loop contribution is dominated by~$m_t^2$, e.g.\ when only coupling the axion to the right-handed top, we deduce $\Lambda_t \gtrsim \SI{4e9}{GeV}$. This corresponds to a contribution of $\Delta\Neff \approx 0.005$ which is much smaller than the high-temperature thermal freeze-out contribution of~$0.027$ or near-term cosmological bounds on~$\Delta\Neff$.

\paragraph{Coupling to Bottom Quarks.}
In the case of an independent axion-bottom interaction, we again follow~\cite{Feng:1997tn} and compute the same loop-induced constraint as described for the interactions with tau~leptons and top~quarks: $\Lambda_b \gtrsim \SI{2.0e6}{GeV}$. Here, we assumed that the axion contribution to the electron coupling is dominated by~$m_b^2$, which is the case if the axion only couples to the right-handed bottom, for instance. Given our conservative estimate of~$\Delta\Neff(\Lambda_b)$, we require a cosmological measurement that excludes $\Delta\Neff \gtrsim 0.048$ to improve upon this bound, but we refer to Appendix~\ref{app:rateComparisonsUncertainties} for a discussion on the uncertainties of this estimate and potential implications for~SO and~\mbox{CMB-S4}.

\paragraph{Coupling to Charm Quarks.}
For diagonal couplings to charm~quarks, we follow the same argument while assuming that the loop contribution is dominated by~$m_c^2$, e.g.\ by only coupling to the right-handed charm. In this way, we infer $\Lambda_c \gtrsim \SI{1.4e5}{GeV}$. When comparing this bound to the predictions based on our conservative estimate of~$\Delta\Neff(\Lambda_c)$, an improvement of this bound requires the exclusion of $\Delta\Neff \gtrsim 0.048$. However, lattice QCD~calculations may reveal that weaker bounds on~$\Delta\Neff$ lead to the same bounds on the interaction strength with charm~quarks, with some constraining power not only accessible for~\mbox{CMB-S4}, but potentially even for the Simons Observatory~(cf.~Appendix~\ref{app:rateComparisonsUncertainties}).

\section{Conclusions}
\label{sec:conclusions}

The high temperatures and densities of the early universe provide an ideal environment to test fundamental physics. Even for extremely weak couplings, new particles could be efficiently produced, potentially leaving a lasting imprint on cosmological observables. Axions and other pseudo-Nambu-Goldstone bosons provide a particularly compelling target as they are naturally light and would therefore leave a measurable imprint on cosmological observables via the effective number of relativistic species,~$\Neff$.\medskip

\begin{figure}[h!t]%
	\centering%
	\includegraphics{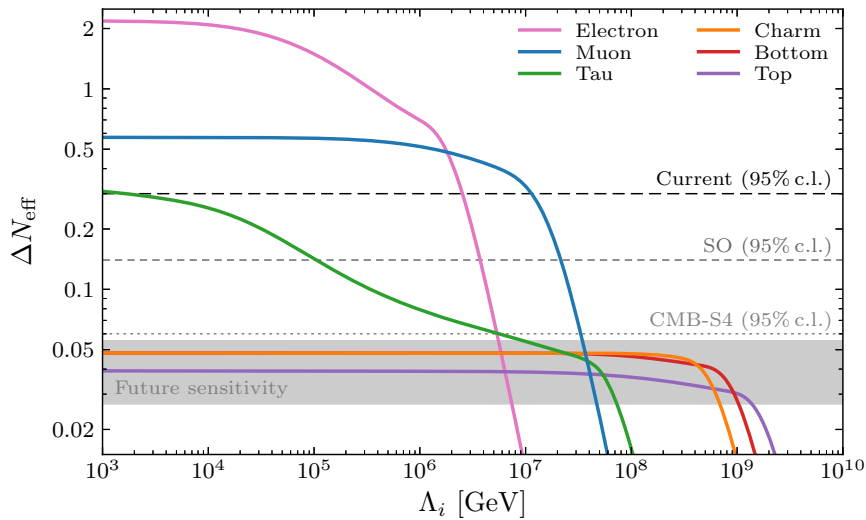}%
	\caption{Contribution to the radiation density as parameterized by~$\Delta\Neff$ as a function of axion coupling strength~$\Lambda_i$ for different Standard Model fermions~$\psi_i$~(cf.~Fig.~\ref{fig:DeltaNeff_lambda_extended}). From this figure, we can translate current and future constraints on~$\Neff$~(cf.~Fig.~\ref{fig:DeltaNeff_freezeout}) into the equivalent bounds on~$\Lambda_i$ for any couplings to matter.}%
	\label{fig:DeltaNeff_summary}%
\end{figure}

In this paper, we calculated the predicted contributions to~$\Neff$ from axions and other~pNGBs that are coupled to Standard Model fermions. We focused on the effectively marginal interactions that arise after electroweak symmetry breaking which can thermalize these particles beyond the Standard Model at low temperature. The axions eventually decouple when the temperature drops below the mass of the associated fermion. Since this happens below the electroweak scale, they contribute $\Delta\Neff > 0.027$ to the radiation density in the early universe which makes them compelling targets for near-term surveys. Our main result is shown in Fig.~\ref{fig:DeltaNeff_summary}, which provides a direct link between the measurement of~$\Neff$ and limits on the coupling to SM~fermions.%
\medskip

This work includes an improved calculation of the thermal axion production rates. The described method for calculating the thermal averages including the full quantum statistics is also relevant to other production rate calculations, including axion production at high temperatures. In addition, the same production rates calculated in this paper also appear in astrophysical constraints on axions. We are therefore able to use the common origin of axion production to compare the strengths and weaknesses of the cosmological and astrophysical bounds on axion couplings to matter, and saw that the cosmological production of axions is essentially determined by dimensional analysis as are the associated bounds.\medskip

Our results are particularly important in the context of ongoing cosmic surveys improving the measurement of~$\Neff$, which are on the precipice of measuring the energy density of a single scalar field decoupling prior to the QCD~phase transition. Concretely, \mbox{CMB-S4}~is expecting to exclude $\Delta\Neff > 0.060$ at 95\%~c.l.~\cite{Abazajian:2019eic} and could be improved in combination with a number of large-scale structure surveys~\cite{Baumann:2017gkg}. The axion couplings to matter discussed in this paper illustrate that this is a particularly compelling level of sensitivity. As several Standard Model fermions have masses around~\SI{1}{GeV}, the coupling of axions to these fermions naturally contributes to~$\Delta\Neff$ at this level. Furthermore, the decoupling temperature depends both on the interaction strength and the mass of the fermion, resulting in a range of~$\Delta\Neff$, even when the axion reaches thermal equilibrium. As a result, there are both numerous thresholds of~$\Delta\Neff$ at this sensitivity and opportunities to continuously improve our understanding of Standard Model couplings of axions and other pseudo-Nambu-Goldstone bosons with the depth of these surveys.

\vskip20pt
\paragraph{Acknowledgments}
The authors thank Nathaniel Craig, Peizhi Du, Peter Graham, Marilena LoVerde, Gustavo Marques-Tavares, Joel Meyers and Surjeet Rajendran for helpful discussions. The authors were supported by the US~Department of Energy under Grants~\mbox{DE-SC0009919} and~\mbox{DE-SC0019035}. B.\,W.\ also acknowledges support from the Simons~Foundation Modern Inflationary Cosmology Initiative under Grant~SFARI~560536. The completion of this work by~B.\,W. was partially supported by a grant from the Simons~Foundation and the hospitality of the Aspen Center for Physics, which is supported by National Science Foundation Grant~\mbox{PHY-1607611}. We acknowledge the use of \texttt{FeynMP}~\cite{Ohl:1995kr}, \texttt{IPython}~\cite{Perez:2007ipy} and~\texttt{RunDec}~\cite{Herren:2017osy}, and the Python packages \texttt{Matplotlib}~\cite{Hunter:2007mat}, \texttt{Numba}~\cite{Lam:2015num}, \texttt{NumPy}~\cite{Harris:2020xlr} and~\texttt{SciPy}~\cite{Virtanen:2019joe}.

\clearpage
\appendix
\section{Computational Details}
\label{app:computationalDetails}

In this appendix, we provide details underlying our calculation of the axion production rate~\eqref{eq:productionRate} that does not rely on approximations of its integrand. While the final integral~\eqref{eq:productionRate2} is eventually evaluated numerically, the analytic reduction of the integral described in~Appendix~\ref{app:computational_rate} is important for making the numerical evaluation tractable. In~Appendix~\ref{app:computational_Boltzmann}, we present additional information on our subsequent calculation of the contributions to~$\Delta\Neff$ using the Boltzmann equation.

\subsection{Production Rate Calculation}
\label{app:computational_rate}

We want to directly compute the production rate of axions via Compton-like scattering and fermion annihilation using the quantum distribution functions~\eqref{eq:quantumDistributions} for the incoming and outgoing particles which we label by~$1,2$ and~$3,4$, respectively. The general interaction rate via such two-to-two processes is given by~\eqref{eq:productionRate}:
\begin{equation}
	\Gamma_\phi = \frac{1}{n^\mathrm{eq}_\phi} \int\! \d\tilde{\Gamma}\, f_1(p_1)\, f_2(p_2) \left[1 \pm f_3(p_3)\right] \left[1 \pm f_4(p_4)\right]\, \sum |\mathcal{M}|^2\, ,	\label{eq:productionRate3}
\end{equation}
where we introduced the measure
\begin{equation}
	\d\tilde{\Gamma} = \prod_{i=1}^{4} \frac{\d^3 p_i}{(2\pi)^3 2 E_i}\, (2\pi)^4 \delta^{(3)} (\p_1 + \p_2 - \p_3 - \p_4) \, \delta(E_1 + E_2 - E_3 - E_4)\, .	\label{eq:measure}
\end{equation}
In the following, we will change the variables of this 12-dimensional integral, rewrite the integration measure and scattering amplitudes~\eqref{eq:ComptonAmplitude} and~\eqref{eq:annihilationAmplitude}, and finally reduce the respective production rates to four-dimensional integrals that we will solve numerically.

\subsubsection*{Parametrization}
While there are 12~degrees of freedom in the integration variables~$\p_i$, $i=1,\ldots,4$, the four-dimensional energy-momentum conservation enforced by the Dirac delta functions reduces the number of independent degrees of freedom to eight. To parametrize these, we employ the following variables: the absolute values of the total energy, total momentum, one incoming momentum and one outgoing momentum,
\begin{equation}
	E = E_1 + E_2\, ,	\qquad p = |\p| = |\p_1 + \p_2|\, ,	\qquad p_1 = |\p_1|\, ,	\qquad p_3 = |\p_3|\, ,
\end{equation}
the polar angle~$\phi_1$ of~$\p_1$, the polar angle~$\phi_3$ of~$\p_3$, and the remaining two directions of~$\p$. Due to symmetry, we can set $\p \equiv p \,\zhat$, $\phi_1 \equiv 0$ and $\phi_3 \equiv \phi$. Given this parametrization, we have $p_2 = |\p_2| = |\p - \p_1|$ and $p_4 = |\p_4| = |\p - \p_3|$.

It is useful to define the angle between~$\p$ and~$\p_i$, $\theta_i = \angle(\p,\p_i)$, to transform the three-momenta of the particles from Cartesian to spherical coordinates:
\begin{equation}
	\begin{split}
		\p_1 = p_1\,(\sin\theta_1, 0, \cos\theta_1)\, , 							\quad&\quad	\p_2 = p_2\,(-\sin\theta_2, 0, \cos\theta_2)\, ,	\\
		\p_3 = p_3\,(\sin\theta_3 \cos\phi, \sin\theta_3\sin\phi, \cos\theta_3)\, , \quad&\quad	\p_4 = p_4(-\sin\theta_4 \cos\phi, -\sin\theta_4 \sin\phi, \cos\theta_4)\, .
	\end{split}
\end{equation}
We can express these angles in terms of the absolute values of the momenta for $i=1,3$ and $j=2,4$ as follows:
\begin{equation*}
	\cos\theta_i = \frac{p^2 + p_i^2 - p_{i+1}^2}{2 p p_i}\, ,\	\sin\theta_i = \frac{g(p,p_i,p_{i+1})}{2p p_i}\, ,\	\cos\theta_j = \frac{p^2 + p_j^2 - p_{j-1}^2}{2 p p_j}\, ,\	\sin\theta_j = \frac{g(p,p_{j-1},p_j)}{2p p_j}\, ,
\end{equation*}
where we introduced $g(p,p_i,p_j) = \sqrt{(p+p_i+p_j)(p+p_i-p_j)(p-p_i+p_j)(-p+p_i+p_j)}$. Finally, it is also helpful to introduce the angles $\theta_{13} = \angle(\p_1, \p_3)$ and $\theta_{14} = \angle(\p_1, \p_4)$, which can be parametrized as
\begin{equation}
\begin{split}
	\cos\theta_{13}	&= \hphantom{-}\sin\theta_1 \sin\theta_3 \cos\phi + \cos\theta_1 \cos\theta_3	\equiv \hphantom{-}a \cos\phi + b\, ,	\\
	\cos\theta_{14}	&= -\sin\theta_1 \sin\theta_4 \cos\phi + \cos\theta_1 \cos\theta_4				\equiv -c \cos\phi + d\, ,
\end{split}	\label{eq:short_abcd}
\end{equation}
where we defined a short-hand notation in terms of~$a$, $b$, $c$ and~$d$ in the last equalities, respectively.

\subsubsection*{Integral Measure}
We now turn to the measure of the production rate integral introduced in~\eqref{eq:measure} and express it in terms of the new coordinates. By inserting $1 = \int\!\d^3p\, \delta^{(3)}(\p - \p_1 - \p_2)$, we get
\begin{align}
	\int\! \d\tilde{\Gamma}	&= \int\!\d^3p_1 \int\!\d^3p_2 \int\!\d^3p_3 \int\!\d^3p_4\, \frac{\delta^{(3)}(\p_1 + \p_2 - \p_3 - \p_4)\, \delta(E_1 + E_2 - E_3 - E_4)}{(2\pi)^8 16 E_1 E_2 E_3 E_4}	\nonumber	\\
							&= \int\!\d^3p_1 \int\!\d^3p_2 \left[\int\!\d^3p\, \delta^{(3)}(\p - \p_1 - \p_2)\right]	\nonumber	\\
	 						& \hphantom{= \int\!\d^3p_1 \int\!\d^3p_2}\, \times \int\!\d^3p_3 \int\!\d^3p_4\, \frac{\delta^{(3)}(\p_1 + \p_2 - \p_3 - \p_4)\, \delta(E_1 + E_2 - E_3 - E_4)}{(2\pi)^8 16 E_1 E_2 E_3 E_4}\\
	 						&= \left. \int\!\d^3p \int\!\d^3p_1 \int\!\d^3p_3\, \frac{\delta(E_1 + E_2 - E_3 - E_4)}{(2\pi)^8 16 E_1 E_2 E_3 E_4} \right|_{\p_2=\p-\p_1\hskip-1pt,\, \p_4=\p-\p_3} ,	\nonumber
\end{align}
where we introduced a short-hand notation for imposing $\p_j = \p - \p_{j-1}$ in the third line which we will further abbreviate below. As previously mentioned, we have the freedom to take $\p = p\, \zhat$ for fixed~$p$ since the direction of~$\p$ does not affect the production rate. Similarly, we can rotate all the particle momenta~$\p_i$ around the total momentum~$\p$ because only their relative angle matters. This motivates choosing the polar angle between the planes spanned by~$(\p_1, \p_2)$ and~$(\p_3, \p_4)$, $\phi = \phi_3 - \phi_1$, as an integration variable (which is equivalent to the choice mentioned above). With these choices, the measure becomes
\begin{align}
	\int\! \d\tilde{\Gamma}	&= \left. \int\!\d p\,p^2 \int\!\d p_1\, p_1^2\, \d\phi_1\, \d\!\cos\theta_1 \int\!\d p_3\, p_3^2\, \d\phi_3\, \d\!\cos\theta_3\, \frac{\delta(E_1 + E_2 - E_3 - E_4)}{\hskip1pt(2\pi)^7 8 E_1 E_2 E_3 E_4} \right|_{\p_2,\, \p_4}	\\
							&= \left. \int\!\d E \int\!\d p\,p^2 \int\!\d\phi \Bigg[\prod_{i=1,3} \int\!\d p_i\,p_i^2 \int\!\d\!\cos\theta_i\, \delta(E - E_i - E_{i+1}) \Bigg] \frac{1}{512\pi^6 E_1 E_2 E_3 E_4} \right|_{\p_2,\, \p_4} ,	\nonumber
\end{align}
where we inserted $1 = \int\!\d E\, \delta(E - E_1 - E_2)$ to introduce the total energy~$E$. To evaluate the integrals over the azimuthal angles~$\theta_i$, we use $E_{i+1} = E - E_i = (m_{i+1}^2 + p^2 + p_i^2 - 2 p p_i \cos\theta_i)^{1/2}$ and $\delta(g(p)) = \sum_k \delta(p-\bar{p}_k)/|g'(\bar{p}_k)|$, with the simple zeros~$\bar{p}_k$ of the function~$g(p)$, to reparametrize the Dirac delta functions:
\begin{equation}
	\int_{-1}^1\!\d\!\cos\theta_i\, \delta(E - E_i - E_{i+1}) = \frac{E_{i+1}}{p p_i} \int_{-1}^1\!\d\!\cos\theta_i\, \delta\!\left(\!\cos\theta_i - \frac{m_{i+1}^2 + p^2 + p_i^2 - E_{i+1}^2}{2 p p_i}\right) ,	\label{eq:cosThetaIntegral}
\end{equation}
which we can directly evaluate subject to the finite integration limits.

Finally, we have to appropriately treat the integration limits which we have neglected so far. The integral in~\eqref{eq:cosThetaIntegral} is non-vanishing only if
\begin{equation}
	\left(\frac{m_{i+1}^2 + p^2 + p_i^2 - E_{i+1}^2}{2 p p_i}\right)^{\!2} \leq 1\, .
\end{equation}
After defining $A = s + m_i^2 - m_{i+1}^2$ and $B = A^2 + 4 E^2 m_i^2$, and introducing the Mandelstam variable $s = E^2 - p^2$, we can expand this inequality to
\begin{equation}
	16s^2\! \left(p_i^2\right)^{\!2} + \left(8sB - 16E^2 A^2\right) p_i^2 + B^2 - 16E^2 A^2 m_i^2 \leq 0\, , \label{eq:inequality}
\end{equation}
which determines the integration limits. To satisfy this quadratic inequality in~$p_i^2$, its determinant must be positive, which requires $s \geq (m_i + m_{i+1})^2$. Since the total momentum~$p$ is positive, this implies for the total momentum and energy:
\begin{equation}
	p \leq \sqrt{E^2 - (m_i + m_{i+1})^2}\, ,	\qquad	E \geq m_i + m_{i+1}\, .
\end{equation}
Because these requirements have to be simultaneously satisfied for $i=1,3$, we get
\begin{equation}
	E_\mathrm{min} = \max_{i=1,3}\left\{m_i+m_{i+1}\right\} ,	\qquad	p_\mathrm{max} = \min_{i=1,3}\!\left\{\!\sqrt{E^2 - ({m_i+m_{i+1})^2}}\right\} ,	\label{eq:limit_E+p}
\end{equation}
which can also be directly inferred from kinematic considerations. The two solutions to the quadratic inequality~\eqref{eq:inequality} for~$p_i^2$ then implies the integration limits for~$p_i$, $i=1,3$, to be
\begin{equation}
	p_i^\mathrm{min, max} = \frac{1}{2s}\left|E\sqrt{[s-(m_i-m_{i+1})^2][s-(m_i+m_{i+1})^2]} \mp (s+m_i^2-m_{i+1}^2)p\right| ,	\label{eq:limit_pi}
\end{equation}
with the minus (plus) sign being associated with the lower (upper) limit. To put it all together, we therefore arrive at the following result for the measure:
\begin{equation}
	\int\! \d\tilde{\Gamma}	= \left. \int_{E_\mathrm{min}}^\infty\! \d E \int_0^{p_\mathrm{max}}\! \d p \int_{p_1^\mathrm{min}}^{p_1^\mathrm{max}}\! \d p_1 \int_{p_3^\mathrm{min}}^{p_3^\mathrm{max}}\! \d p_3 \int_0^{2\pi}\! \d\phi\,\frac{p_1 p_3}{512\pi^6 E_1 E_3} \right|_{\p_2,\,\p_4} ,
\end{equation}
with the integration limits given by~\eqref{eq:limit_E+p} and~\eqref{eq:limit_pi}.

\subsubsection*{Scattering Amplitudes}
Before turning to the entire production rate calculation, we first rewrite the amplitudes of Compton-like scattering and fermion annihilation, which we provided in~\eqref{eq:ComptonAmplitude} and~\eqref{eq:annihilationAmplitude}, as a function of the invariant Mandelstam variables~$s$, $t$ and~$u$. In terms of our integration variables, these invariants are given by
\begin{align}
	s &= (E_1 + E_2)^2 - (\p_1 + \p_2)^2 = E^2 - p^2\, ,												\nonumber	\\
	t &= (E_1 - E_3)^2 - (\p_1 - \p_3)^2 = m_1^2 + m_3^2 - 2 E_1 E_3 + 2 p_1 p_3 \cos\theta_{13}\, ,				\\
	u &= (E_1 - E_4)^2 - (\p_1 - \p_4)^2 = m_1^2 + m_4^2 - 2 E_1 E_4 + 2 p_1 p_4 \cos\theta_{14}\, .	\nonumber
\end{align}
This implies that the scattering amplitude of the Compton-like process can be rewritten as
\begin{equation}
	\sum |\M|^2_{(a)} = 16\pi\hskip1pt A_\psi\, |\tilde \epsilon_\psi|^2 \frac{2 p_1 p_3^2 (1 - \cos\theta_{13})^2}{(E^2 - p^2 - m_\psi^2) \left(E_4 - p_4 \cos\theta_{14}\right)}\, ,
\end{equation}
while the scattering amplitude of the annihilation process in these coordinates is
\begin{equation}
	\sum |\M|^2_{(b)} = 16\pi\hskip1pt A_\psi\, |\tilde \epsilon_\psi|^2 \frac{(E^2 - p^2)^2}{4 p_3 p_4(E_1 - p_1\cos\theta_{13})(E_1 - p_1\cos\theta_{14})}\, .
\end{equation}
We will now separately insert these expressions into the integral~\eqref{eq:productionRate3} to obtain the final expressions for the respective production rates.

\subsubsection*{Compton-Like Scattering Rate}
For Compton-like scattering, $\{\gamma,g\} + \psi \to \phi + \psi$, we take the momenta of the massless bosons to be~$p_1$ and~$p_3$, while the fermion~$\psi$ has incoming momentum~$p_2$, outgoing momentum~$p_4$ and mass~$m_\psi$. This means that the energies and momenta can be expressed as
\begin{equation*}
	E_1 = p_1\, ,	\qquad	E_3 = p_3\, ,	\qquad	E_j = E - p_{j-1}\, ,	\qquad	p_j = \sqrt{(E - p_{j-1})^2 - m_\psi^2}\, ,
\end{equation*}
for $j=2,4$, while the integration limits for $i=1,3$ are
\begin{equation*}
	E_\mathrm{min} = m_\psi\, ,	\quad	p_\mathrm{max} = \sqrt{E^2 - m_\psi^2}\, ,	\quad	p_i^\mathrm{min} = \frac{E^2-m_\psi^2-p^2}{2(E+p)}\, ,	\quad	p_i^\mathrm{max} = \frac{E^2-m_\psi^2-p^2}{2(E-p)}\, .
\end{equation*}
The production rate associated with the scattering process therefore is
\begin{align*}
	\Gamma_{(a)}	&= \frac{1}{512\pi^6\, n^\mathrm{eq}_\phi} \int\!\d E \int\!\d p \int\!\d p_1\! \int\!\d p_3\! \int\!\d\phi\, f_1(E_1)\, f_2(E_2) \left[1 + f_3(E_3)\right] \left[1 - f_4(E_4)\right])\, \sum|\mathcal{M}|^2	\\
					&= \frac{A_\psi |\tilde{\epsilon}_\psi|^2}{16\pi^5\, n^\mathrm{eq}_\phi} \int\!\d E \int\!\d p \int\!\d p_1\! \int\!\d p_3\, \frac{1}{\left(\ee^\frac{p_1}{T}-1\right)\! \left(\ee^\frac{E-p_1}{T}+1\right)\! \left(1-\ee^{-\frac{p_3}{T}}\right)\! \left(1+\ee^{-\frac{E-p_3}{T}}\right)}	\\
					&\hphantom{= \frac{A_\psi |\tilde{\epsilon}_\psi|^2}{16\pi^5\, n^\mathrm{eq}_\phi} \int\!\d E \int\!\d p \int\!\d p_1 \int\!\d p_3}\, \times \frac{p_1 p_3^2}{E^2 - p^2 - m_\psi^2} \int_0^{2\pi}\!\d\phi\, \frac{(1 - \cos\theta_{13})^2}{E_4 - p_4\cos\theta_{14}}\, .
\end{align*}
We can further simplify this expression by analytically performing the angular integral,
\begin{equation*}
	\frac{1}{2\pi} \int_0^{2\pi}\!\d\phi\, \frac{p_4 (1-\cos\theta_{13})^2}{E_4 - p_4\cos\theta_{14}} = \frac{c^2(1-b)^2 + [a^2f+2ac(1-b)](f-\sqrt{f^2-c^2})}{c^2\sqrt{f^2-c^2}}\, ,
\end{equation*}
where we employed the short-hand notation introduced in~\eqref{eq:short_abcd}, with
\begin{equation*}
	\cos\theta_i = \frac{m_\psi^2-E^2+p^2+2E p_i}{2p p_i}\, ,	\hspace{12pt}	\cos\theta_4 = \frac{E^2-m_\psi^2+p^2-2E p_3}{2p \sqrt{(E-p_3)^2-m_\psi^2}}\, ,	\hspace{12pt}	f = \frac{E - p_3}{\sqrt{(E-p_3)^2-m_\psi^2}} - d\, ,
\end{equation*}
for $i=1,3$. Finally, we can use $a/c = p_4/p_3$ to arrive at the final expression for the Compton-like production rate,
\begin{equation*}
	\begin{split}
		\Gamma_{(a)}	&= \frac{A_\psi |\tilde{\epsilon}_\psi|^2}{8\pi^4\,n^\mathrm{eq}_\phi} \int_{m_\psi}^\infty\!\d E \int_0^{p_\mathrm{max}}\!\d p \int_{p_1^\mathrm{min}}^{p_1^\mathrm{max}}\!\d p_1 \int_{p_3^\mathrm{min}}^{p_3^\mathrm{max}}\!\d p_3\, \frac{1}{\left(\ee^\frac{p_1}{T}-1\right)\!\left(\ee^\frac{E-p_1}{T}+1\right)}\,\frac{p_1}{E^2 - p^2 - m_\psi^2} \\
						&\hspace{58pt} \times \frac{1}{\left(1-\ee^{-\frac{p_3}{T}}\right)\!\left(1+\ee^{-\frac{E-p_3}{T}}\right)}\, \frac{p_3^2(1-b)^2 + \left[p_4^2f+2p_3p_4(1-b)\right]\left(f-\sqrt{f^2-c^2}\right)}{p_4\sqrt{f^2-c^2}}\, ,
	\end{split}
\end{equation*}
with $p_4^2\,(f^2-c^2) = (E_4 \cos\theta_1 - p_4 \cos\theta_4)^2 + m_\psi^2 \sin^2\theta_1$.

\subsubsection*{Fermion Annihilation Rate}
For fermion-antifermion annihilation, $\psi + \bar{\psi} \to \{\gamma,g\} + \phi$, we assign the incoming momenta~$p_1$ and~$p_2$ to the fermion and antifermion with mass~$m_\psi$, and the outgoing momenta~$p_3$ and~$p_4$ to the massless vector boson and axion, respectively. In consequence, the energies and momenta associated with this process are
\begin{equation*}
	E_1 = \sqrt{p_1^2 + m_\psi^2}\, ,	\quad	E_3 = p_3\, ,	\quad	E_j = E - p_{j-1}\, ,	\quad	p_2 = \sqrt{(E - E_1)^2-m_\psi^2}\, ,	\quad	p_4 = E - p_3\, ,
\end{equation*}
for $j=2,4$, while the integration limits are given by
\begin{equation*}
	E_\mathrm{min} = 2 m_\psi\, ,	\quad	p_\mathrm{max} = \sqrt{E^2 - 4 m_\psi^2}\, ,	\quad	p_1^{\mathrm{min},\mathrm{max}} = \frac{1}{2}\left|\frac{E}{\sqrt{s}}\sqrt{s-4m_\psi^2} \mp p\right|\, ,	\quad	p_3^{\mathrm{min},\mathrm{max}} = \frac{E \mp p}{2}\, ,
\end{equation*}
where the minus (plus) signs are associated with the minimum (maximum) particle momenta. We can therefore write the interaction rate for the annihilation process as
\begin{align*}
	\Gamma_{(b)}	&= \frac{1}{512\pi^6\, n^\mathrm{eq}_\phi} \int\!\d E \int\!\d p \int\!\d p_1\! \int\!\d p_3\! \int\!\d\phi\, \frac{p_1}{E_1}\hskip1pt f_1(E_1)\hskip1pt f_2(E_2)\! \left[1 + f_3(E_3)\right]\! \left[1 + f_4(E_4)\right]) \sum|\mathcal{M}|^2	\\
					&= \frac{A_\psi |\tilde{\epsilon}_\psi|^2}{128\pi^5\, n^\mathrm{eq}_\phi} \int\!\d E \int\!\d p \int\!\d p_1\! \int\!\d p_3\, \frac{1}{\left(\ee^\frac{E_1}{T}+1\right)\! \left(\ee^\frac{E-E_1}{T}+1\right)\! \left(1-\ee^{-\frac{p_3}{T}}\right)\! \left(1-\ee^{-\frac{E-p_3}{T}}\right)}	\\
					&\hphantom{= \frac{A_\psi |\tilde{\epsilon}_\psi|^2}{128\pi^5\, n^\mathrm{eq}_\phi} \int\!\d E \int\!\d p \int\!\d p_1 \int\!\d p_3}\, \times \frac{(E^2 - p^2)^2}{E_1 p_3 p_4} \int_0^{2\pi}\!\d\phi\, \frac{p_1}{(E_1 - p_1\cos\theta_{13})(E_1 - p_1\cos\theta_{14})}\, .
\end{align*}
As for the Compton-like process, we can again compute the angular integral analytically and express it in terms of the short-hand notation introduced in~\eqref{eq:short_abcd},
\begin{equation*}
	\frac{1}{2\pi}\int_0^{2\pi}\!\d\phi\, \frac{p_1^2}{(E_1 - p_1\cos\theta_{13})(E_1 - p_1\cos\theta_{14})} = \frac{\frac{a}{\sqrt{(f-b)^2-a^2}} + \frac{c}{\sqrt{(f-d)^2-c^2}}}{a(f-d)+c(f-b)}\, ,
\end{equation*}
with $f = E_1/p_1$ and the following underlying expressions:
\begin{equation*}
	\cos\theta_1 = \frac{m_\psi^2-(E-E_1)^2+p^2+p_1^2}{2p p_1}\, ,	\quad	\cos\theta_3 = \frac{-E^2+p^2+2E p_3}{2p p_3}\, ,	\quad	\cos\theta_4 = \frac{E^2+p^2-2E p_3}{2p (E-p_3)}\, .
\end{equation*}
After rewriting the result of the angular integral, the final expression for the production rate via fermion annihilation that we implemented numerically is given by
\begin{equation*}
	\begin{split}
		\Gamma_{(b)}	&= \frac{A_\psi |\tilde{\epsilon}_\psi|^2}{32\pi^4\,n^\mathrm{eq}_\phi} \int_{2m_\psi}^\infty\!\d E \int_0^{p_\mathrm{max}}\!\d p \int_{p_1^\mathrm{min}}^{p_1^\mathrm{max}}\!\d p_1 \int_{p_3^\mathrm{min}}^{p_3^\mathrm{max}}\!\d p_3\, \frac{1}{\left(\ee^\frac{E_1}{T}+1\right)\!\left(\ee^\frac{E-E_1}{T}+1\right)}\,\frac{(E^2 - p^2) p_1}{E_1} \\
						&\hspace{191pt} \times \frac{1}{\left(1-\ee^{-\frac{p_3}{T}}\right)\!\left(1-\ee^{-\frac{E-p_3}{T}}\right)} \left(\frac{1}{p_3\, h(\theta_3)} + \frac{1}{p_4\, h(\theta_4)}\right) ,
	\end{split}
\end{equation*}
where we defined $h(\theta_k) = \sqrt{m_\psi^2 \sin^2\theta_1 + (E_1\cos\theta_1 - p_1\cos\theta_k)^2}$.

\subsection[Boltzmann Equation and \texorpdfstring{$\Delta\Neff$}{Delta Neff}]{Boltzmann Equation and $\Delta\mathbf{N}_\mathbf{eff}$}
\label{app:computational_Boltzmann}

Having obtained the total production rate $\Gamma = 2\Gamma_{(a)} + \Gamma_{(b)}$ by numerically computing the four-dimensional integrals stated above, we computed the resulting contribution to the radiation density in the early universe as parameterized by~$\Neff$. In the following, we provide additional details of the underlying computational steps.\medskip

First, we solve the Boltzmann equation~\eqref{eq:BoltzmannEquation} to calculate the axion number density~$n_\phi(t)$. Instead of directly solving~\eqref{eq:BoltzmannEquation}, we however adopt the conventional change of variables to the dimensionless time variable $x = m/T$ and the dimensionless comoving number density $Y_\phi = n_\phi/s$, with the entropy density $s = 2\pi^2\, g_{*s} T^3/45$. Conservation of entropy in the early universe, $a^3 s = \mathrm{const}$, implies $\dot{s}/s = -3\hskip1pt \dot{a}/a = -3 H$, where the overdot denotes a derivative with respect to time~$t$, and allows to express~\eqref{eq:BoltzmannEquation} as
\begin{equation}
	\dot{Y}_\phi = \Gamma_\phi\, \big(Y_\phi^\mathrm{eq} - Y_\phi\big)\, .
\end{equation}
Rewriting the derivative with respect to~$t$ in terms of~$x$ leads to
\begin{equation}
	H x \frac{\d Y_\phi}{\d x} = \left(1 - \frac{1}{3} \frac{\d\log g_{*s}}{\d\log x}\right) \Gamma_\phi\, \big(Y_\phi^\mathrm{eq} - Y_\phi\big)\, ,
\end{equation}
where we used the definition of the entropy density~$s$.

We numerically solve this equation for~$Y_\phi$ from an initial condition of no axions, \mbox{$Y_\phi(T_0) = 0$}, to the final late-time value of $Y_{\phi,\infty} = Y_\phi(T_\infty)$. We take the initial temperature to be the temperature of the electroweak crossover, $T_0 = T_\mathrm{EW} = \SI{159.5}{GeV}$~\cite{DOnofrio:2015gop}, since the Lagrangian~\eqref{eq:lagrangian2} and, therefore, the computed production rates are only valid after electroweak symmetry breaking. Due to Boltzmann suppression for $T \ll m_\psi$, it is sufficient to compute the number density for $T_\infty = m_\psi/100$.\medskip

Finally, we have to convert the computed value of~$Y_{\phi,\infty}$ to a contribution to~$\Delta\Neff$ as defined in~\eqref{eq:DeltaNeff}. At late times, the axion energy and number densities, the photon density and the entropy density are given by
\begin{equation}
	\rho_\phi = \frac{\pi^2}{30} T_\phi^4\, ,	\qquad	n_\phi = \frac{\zeta(3)}{\pi^2} T_\phi^3\, , \qquad \rho_\gamma = \frac{\pi^2}{15} T_\gamma^4\, ,	\qquad	s = \frac{2\pi^2}{45} g_{*s,\infty} T_\gamma^3\, ,
\end{equation}
with the effective number of relativistic degrees of freedom in entropy only counting photons and neutrinos, $g_{*s,\infty} = 43/11$. We therefore arrive at the following expression for the contribution to the effective number of relativistic degrees of freedom:
\begin{equation}
	\Delta\Neff = \frac{4}{7} \left(\frac{11\pi^4}{90 \zeta(3)}g_{*s,\infty}\,Y_{\phi,\infty}\right)^{\!4/3}\! = \frac{4}{7} \left(\frac{43\pi^4}{90 \zeta(3)}\,Y_{\phi,\infty}\right)^{\!4/3} ,
\end{equation}
which approximately evaluates to $\Delta\Neff \approx 74.84\, Y_{\phi,\infty}^{4/3}$.

\clearpage
\section{Production Rate Comparisons and Uncertainties}
\label{app:rateComparisonsUncertainties}

In this appendix, we examine a few aspects of the computed production rate of axions and other~pNGBs. We first consider the differences between employing the full quantum distribution functions~$f^i(p)$, $i=f,b$, instead of the classical Boltzmann distribution function and the relative importance of the Compton-like and fermion-antifermion annihilation processes in the production~(Appendix~\ref{app:rate_comparison}). Then, we describe the uncertainties associated with our calculations involving the axion coupling to the bottom and charm~quarks due to the QCD~phase transition and their potential impact on our predictions for~$\Delta\Neff$ as a function of the coupling constants~$\Lambda_{\{b,c\}}$~(Appendix~\ref{app:rate_uncertainties}).

\subsection{Quantum Statistics and Production Rates}
\label{app:rate_comparison}

Commonly, the distribution functions in the production rate are approximated by Boltzmann distributions and the Bose enhancement and Pauli blocking are neglected. We went beyond these approximations and consistently included the quantum nature using the Bose-Einstein and Fermi-Dirac distribution functions, including the effects of Bose enhancement and Pauli blocking.\medskip

When neglecting the quantum statistics, it is convenient to integrate out the momenta of the outgoing particles into the cross-section in the center-of-mass frame. Since the Bose-enhancement and Pauli-blocking terms depend on the energy of the outgoing particles, they however render the cross-section integral more complicated. In this case, some previously employed approximation schemes break down because the approximate integrand peaks in unphysical regimes. Unlike for freeze-out calculations above the electroweak scale, this is particularly noticeable for axion couplings to matter at lower temperatures, as considered in this work. While the calculation of~\cite{Baumann:2016wac} included Bose enhancement and Pauli blocking in a simplified fashion as $\big[1 \pm f_3\big] \big[1 \pm f_4\big] \to \frac{1}{2} \big( [1 \pm f_3(p_1)] [1 \pm f_4(p_2)] + \{ p_1 \leftrightarrow p_2 \} \big)$, this meant that the outgoing momenta were approximated by the incoming momenta. In the case of an incoming and outgoing boson, e.g.\ in the Compton-like process, the integrand however diverges at low momenta,
\begin{equation}
	\lim_{p_1 \to 0} \frac{\d^3 p_1}{2 E_1}\, f^b_1(p_1) \Big[1+f^b_3(p_1)\Big] \sim \frac{\d p_1}{p_1}\, .	\label{eq:low_momentum}
\end{equation}
We remedied these shortcomings by going beyond any of these approximations and incorporating the full quantum statistics in our analytic and numerical computation of the production rate as described in Appendix~\ref{app:computationalDetails}. At the same time, this allows us to compare our full calculation to the results based on the commonly-employed Boltzmann approximation without Bose enhancement and Pauli blocking.

\begin{figure}
	\centering
	\includegraphics{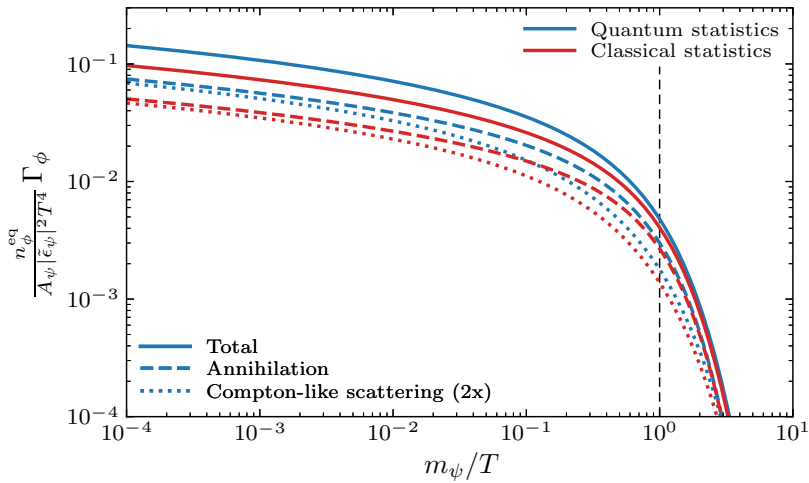}\vspace{-0.05in}
	\caption{Dimensionless rescaling of the total interaction rate~$\Gamma_\phi$ as a function of~$m_\psi/T$ together with its contributions from fermion-antifermion annihilation and Compton-like scattering~(for both fermions and antifermions). We compare the results of our full calculation with the approximate result neglecting the Bose-Einstein and Fermi-Dirac statistics as well as Bose enhancement and Pauli blocking~(see Fig.~\ref{fig:dimensionlessInteractionRate}). The vertical dashed line indicates $T = m_\psi$ which is approximately the temperature where decoupling occurs. We see that the annihilation rate is larger than the Compton-like scattering rate at all temperatures. We also observe that the difference between using the Boltzmann approximation instead of the full quantum statistics is most pronounced for the annihilation rate at large temperatures~$T \gg m_\psi$.}
	\label{fig:dimensionlessInteractionRate_separate}
\end{figure}

We present the results of this comparison of quantum and classical statistics in Fig.~\ref{fig:dimensionlessInteractionRate_separate}%
for the processes relevant to the axion production calculated in this work.\footnote{We note that we assumed production to occur after electroweak symmetry breaking in our calculation of the underlying scattering amplitudes, i.e.\ these results should not be extrapolated to arbitrarily large energies.} We see that the difference between the quantum and classical production rates is fairly substantial when $T \gg m$. This considerable difference justifies the concern that the employed statistics can non-trivially impact production rates and, therefore, the bounds on the couplings from $\Delta\Neff$~measurements. Having said that, the impact on the production rate relevant to these bounds is actually somewhat small since the dominant source of axions will be produced when $T \approx m$, and the effects of Bose enhancement and Pauli blocking are reduced when the number densities are suppressed at low temperatures.

In addition, we separately break down the impact on the annihilation and Compton-like processes. We can observe that the production rate receives approximately equal contributions from Compton-like scattering and fermion-antifermion annihilation at high temperatures, $T \gg m_\psi$, but is dominated by the former and latter process for $T \ll m_\psi$ and around $T = m_\psi$, respectively. In addition, we see that both processes show the same difference between their classical and quantum evaluation for $T \gtrsim m_\psi/10$, with the quantum annihilation rate approaching its classical counterpart for low temperatures whereas the quantum Compton-like scattering rate remains elevated compared to its classical treatment. While this is not directly related to the failure of the approximation scheme for the Compton-like process in~\eqref{eq:low_momentum}, both effects are tied to the correct implementation of the Bose enhancement.

\subsection{Uncertainties in the Axion-Quark Calculation}
\label{app:rate_uncertainties}

The masses of the bottom and charm~quarks are close to the energy scale of the QCD~phase transition. As a result, we expect that decoupling of the axion occurs during the QCD~phase transition for its couplings to these quarks because this happens at temperatures $T \approx m_i /10$ for large interaction strengths. Since the production rates involve external gluons, this would in principle require a non-perturbative calculation, such as with lattice~QCD, to determine the exact contribution to~$\Delta\Neff$. This is a particularly critical issue as the number of degrees of freedom changes rapidly with temperature during the transition, which implies that the predictions of~$\Delta\Neff(\Lambda_{\{b,c\}})$ are extremely sensitive to the temperature of decoupling.\medskip

In the absence of a non-perturbative calculation of the axion production rate,\footnote{Alternatively, one could also attempt to match across the QCD~phase transition, along the lines of~\cite{DEramo:2021psx, DEramo:2021lgb}.} we report our results in terms of conservative and less-conservative estimates for the processes involving the bottom and charm~quarks. In the main text, we only presented the conservative estimates which were computed by cutting off the Boltzmann evolution at a final temperature of $T_\infty = \SI{1}{GeV}$ when the strong coupling constant $\alpha_s \approx 0.5$. In essence, we force the axion decoupling by hand at the onset of the QCD~phase transition when perturbation theory starts to break down. This likely underestimates the contributions~$\Delta\Neff$ at larger couplings~(smaller~$\Lambda_i$) for which the axion is surely still in equilibrium for $T < \SI{1}{GeV}$.

\begin{figure}
	\centering
	\includegraphics{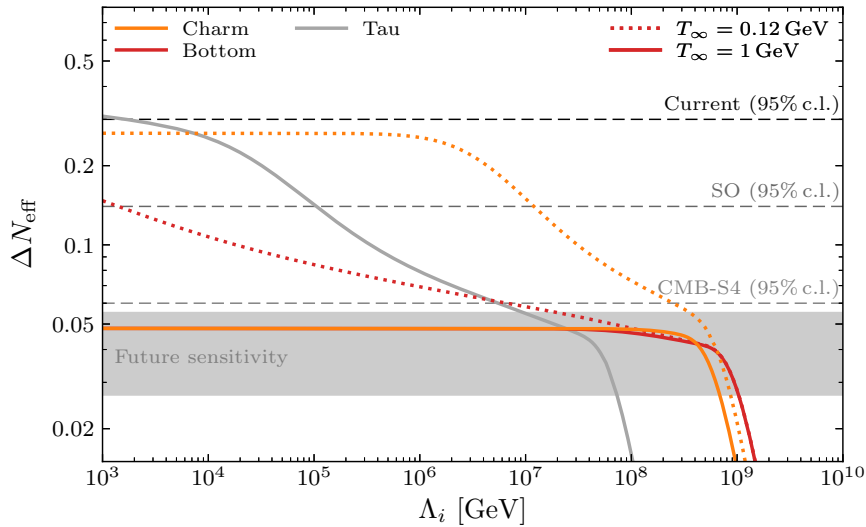}
	\caption{Contributions to~$\Delta\Neff$ from coupling to the charm~(orange) and bottom~(red) quarks. For comparison, the gray line indicates the prediction for axion-tau interactions (cf.~Figures~\ref{fig:DeltaNeff_lambda_extended} and~\ref{fig:DeltaNeff_summary}). The solid line indicates the conservative assumption made in the main text where axion decoupling is imposed by hand at $T_\infty = \SI{1}{GeV}$. The dashed line indicates our less-conservative estimate where we allow the axions to remain in equilibrium through the QCD~phase transition, with $\alpha_s = 1$, and force decoupling at $T_\infty = \SI{120}{MeV}$. We see that the predictions depend sensitively on non-perturbative physics during the QCD~phase transition.}
	\label{fig:DeltaNeff_lambda_uncertainty}
\end{figure}

In Figure~\ref{fig:DeltaNeff_lambda_uncertainty}, %
we included a less conservative calculation~(dashed lines), that stops at the lower end of the QCD~phase transition, $T_\infty = \SI{120}{MeV}$, assuming that the strong coupling~$\alpha_s$ is fixed to unity in the regime where the naive coupling would exceed one. We note that the results of~\cite{Ferreira:2018vjj} show the same qualitative behavior. This figure illustrates that there is little difference between the two estimates for weak couplings~(large~$\Lambda_i$) because the axion is no longer in equilibrium during the QCD~phase transition. The difference becomes large at stronger interaction strengths~(smaller~$\Lambda_i$) since the axion remains in equilibrium at $T < \SI{1}{GeV}$ in the less-conservative scenario. Larger couplings keep the axion in equilibrium to progressively lower temperatures and the contribution to~$\Delta\Neff$ climbs accordingly. The less conservative estimate suggests that the coupling to the bottom and charm~quarks may lie within the sensitivity of~\mbox{CMB-S4} and potentially even the Simons Observatory. This motivates a non-perturbative computation of axion production from bottom and charm couplings during the QCD~phase transition.

\clearpage
\phantomsection
\addcontentsline{toc}{section}{References}
\bibliographystyle{utphys}
\bibliography{references}

\end{document}